\documentclass[nofootinbib,amsmath,amssymb,aps,pre,showkeys,onecolumn,eqsecnum,notitlepage]{revtex4-1}
\usepackage{graphicx}
\usepackage{bm}

\usepackage[varg]{txfonts}
\usepackage{graphics}
\usepackage[utf8]{inputenc}

\begin{document}
\title{Effects of Turbulent Environment and Random Noise on Self-Organized Critical Behavior: \\
Universality vs Nonuniversality}

\author{N.~V.~Antonov$^{1}$}
\email{n.antonov@spbu.ru}
\author{N.~M.~Gulitskiy$^{1}$}
\email{n.gulitskiy@spbu.ru}
\author{P.~I.~Kakin$^{1}$}
\email{p.kakin@spbu.ru}
\author{V.~D.~Serov$^{1,2}$}

\affiliation{$^{1}$Department of Physics, Saint Petersburg State University, 7/9~Universitetskaya~nab., Saint~Petersburg 199034, Russian Federation\\
$^{2}$Department of Theoretical Physics, Peter the Great Saint Petersburg Polytechnic University,
29~Polytechnicheskaya~st., Saint Petersburg 195251, Russian Federation}

%%%%%%%%%%%%%%%%%%%%%%%%%%%%%%%%%%%%%%%%%%%%%%%%%%%%%%%%%%%%%%%%%%%%%%%%%%%%%%%%%%%%%%%%%%%%%%%%%%%%%%%%%%%%%%%%%%%%%%%%%%%%%%%%%%%%%%%%%%%%%%%%%%%%%%%%%%%%%%%%%%%%%%%%%%%%%%%%%%%%%%%%%%

\begin{abstract}
Self-organized criticality in the Hwa--Kardar model of ``running sandpile'' [{\it Phys. Rev. Lett.} {\bf 62}, 1813~(1989);
{\it Phys. Rev.} A {\bf 45}, 7002~(1992)] with a turbulent motion of the environment taken into account is studied with the field theoretic renormalization group (RG). The turbulent flow is modelled by the synthetic $d$-dimensional generalization of the anisotropic Gaussian velocity ensemble with finite correlation time,
introduced by Avellaneda and Majda [{\it Commun. Math. Phys.} {\bf 131}, 381~(1990); {\bf 146}, 139 (1992)].
The \mbox{Hwa--Kardar} model with time-independent (spatially quenched) random noise is considered alongside the original model with the white noise. The aim of the present paper is to explore fixed points of the RG equations which determine the possible types of universality classes (regimes of critical behavior of the system) and critical dimensions of the measurable quantities. Our calculations demonstrate that influence of the type of the random noise is extremely large: in contrast to the case of the white noise where the system possess three fixed points, the case of the spatially quenched noise involves four fixed points with overlapping stability regions. This means that in the latter case the critical behavior of the system depends not only on the global parameters of the system which is the usual case, but also on the initial values of the charges (coupling constants) of the system. These initial conditions determine the specific fixed point which will be reached by the RG flow. Since now the critical properties of the system are not defined strictly by its parameters, the situation may be interpreted as universality violation. Such systems are not forbidden but they are rather rare. It is especially interesting that the same model without turbulent motion of the environment does not predict this nonuniversal behavior and demonstrates the usual one with prescribed universality classes instead [J.~Stat. Phys. {\bf 178}, 392 (2020)].
\end{abstract}

\maketitle

%%%%%%%%%%%%%%%%%%%%%%%%%%%%%%%%%%%%%%%%%%%%%%%%%%%%%%%%%%%%%%%%%%%%%%%%%%%%%%%%%%%%%%%%%%%%%%%%%%%%%%%%%%%%%%%%%%%%%%%%%%%%%%%%%%%%%%%%%%%%%%%%%%%%%%%%%%%%%%%%%%%%%%%%%%%%%%%%%%%%%%%%%%

\section{Introduction} \label{sec:Intro}

%%%%%%%%%%%%%%%%%%%%%%%%%%%%%%%%%%%%%%%%%%%%%%%%%%%%%%%%%%%%%%%%%%%%%%%%%%%%%%%%%%%%%%%%%%%%%%%%%%%%%%%%%%%%%%%%%%%%%%%%%%%%%%%%%%%%%%%%%%%%%%%%%%%%%%%%%%%%%%%%%%%%%%%%%%%%%%%%%%%%%%%%%%

Since its introduction, the concept of self-organized criticality (SOC)~\cite{BTW,BTW1,BTW2,Bak,Bak1,Bak2,Bak3} has been a focus of constant attention and scrutiny~\cite{Col0,Col0,Col1,Col2,Col3}. In a stark contrast to equilibrium systems that display critical scaling (long-time and large-distance asymptotic behavior with
universal exponents) when a tuning parameter (e.g., the temperature) approaches a critical value~\cite{Amit,Zinn,Book3}, the systems with SOC arrive at the critical state due to their intrinsic dynamics. This ``self-tuning'' is observed in various open nonequilibrium systems with dissipative transport including biological systems~\cite{bio1,bio2}, their subclass neural systems~\cite{neu1,neu2,neu3,neu4,neu5,neu6}, online social network systems~\cite{net1,net2,net3,net4,net5,net6}, and various others. As advanced data analysis and sophisticated computational methods become more available, researchers from various fields increasingly turn to the concept of SOC. For example, in~\cite{crop} SOC was used to explain connection between crop losses and extreme climate events while in~\cite{autism} crisis behavior in autism spectrum disorders was analyzed as a self-tuned critical state.

SOC is usually described by discrete models, with discrete space and time evolution. For example, in~\cite{net1} the model of a disordered system of interacting spins was used to determine the primary mechanism for self-tuning in a social network for human collaborative knowledge creation. Nevertheless, universal scaling properties of SOC can be studied using simplified continuous models for smoothed (coarse-grained) fields. Indeed, this approach proved to be fruitful for investigation of critical behavior of various discrete systems. For example, it was found that the discrete Ising and Heisenberg models of equilibrium critical behavior belong to the universality class of the continuous $O(n)$-symmetric $\varphi^4$ model; see~\cite{Amit,Zinn,Book3}. A nonequilibrium example is provided by growth phenomena and fluctuating surfaces~\cite{Halpin}, where numerous discrete models are believed to belong to the universality class of the continuous Kardar--Parisi--Zhang model~\cite{FNS,KPZ}; for a recent discussion of reaction-diffusion models see~\cite{WieseSP}. The conserved directed-percolation related to the Manna universality class of SOC is also often studied by continuous models, see recent papers~\cite{SDP1,SDP2}.

In papers~\cite{HK,HK1}, Hwa and Kardar proposed an anisotropic stochastic differential equation as a continuous model for a system with SOC. The equation describes evolution of a sandpile surface that undergoes changes as new sand enters the system and triggers avalanches (``running'' sandpile). The surface has a flat average slope that determines the preferred direction for the sand transport.

Let us describe the model. The stochastic equation for the scalar field
$h(x)=h(t,{\bm x})$ that denotes a deviation of the sand profile height from its average value is taken in the form
\begin{equation}
\partial_{t} h= \nu_{\bot 0}\, \boldsymbol{\partial}_{\bot}^{2} h + \nu_{\parallel 0}\,
\partial_{\parallel}^{2} h -
\partial_{\parallel} h^{2}/2 + f.
\label{eq1}
\end{equation}
A unit constant vector ${\bm n}$ defines the preferred direction, so any vector ${\bm x}$ decomposes as ${\bm x} = {\bm x}_{\bot} + {\bm n}\, x_{\parallel}$ where
$({\bm x}_{\bot} \cdot \,{\bm n}) =0$. This leads to the appearance of the two spatial derivatives: a $(d-1)$-dimensional gradient $\boldsymbol{\partial_{\bot}}$ and one-dimensional \mbox{gradient $\partial_{\parallel}$}.
The former $\boldsymbol{\partial_{\bot}}=\partial/ \partial {x_{i}}$ with $i=1,\dots, (d-1)$ is
the derivative in the subspace orthogonal to ${\bm n}$, the latter is defined as
$\partial_{\parallel} = ({\bm n} \cdot \boldsymbol{\partial})$. Symbol $d$ denotes the spatial dimension,
$\partial_{t} = \partial/ \partial {t}$, $\boldsymbol{\partial}_{\bot}^{2}=(\boldsymbol{\partial}_{\bot}\cdot\boldsymbol{\partial}_{\bot})$, $\nu_{\parallel 0}$ and
$\nu_{\bot 0}$ are two diffusivity coefficients, and $f(x)$ is a random noise. Traditionally, the nonlinear term $\partial_{\parallel} h^{2}/2$ would have a coupling constant as a factor. Here the fields and the parameters were rescaled to make this factor equal to unity (the coupling constant, thus, appears in the amplitude of the correlator for the random noise $f$).

Different types of the random noise correspond to different physical systems and, as we will see, lead to completely different critical properties. A white noise $f_{w}(x)$, i.e., a Gaussian random
noise with zero mean and the pair correlation function of the form
\begin{equation}
\langle f_{w}(x)f_{w}(x') \rangle = C_{0}\,
\delta(t-t')\, \delta^{(d)}({\bm x}-{\bm x}'), \quad
C_0>0,
\label{forceD}
\end{equation}
was used in~\cite{HK,HK1}.
A generalization of the Hwa--Kardar model with this noise and the coupling constant that was also considered to be a random field was proposed and studied in~\cite{Tadic}. A model similar to~\eqref{eq1}~-- (\ref{forceD}) with the nonlinearity $\propto \partial_{\parallel}^2 h^{3}$
was introduced in~\cite{Pastor1,Pastor2} and discussed in relation to erosion of landscapes.\footnote{It should be noted that such a modification leads to drastic changes in the RG analysis: the model~\cite{Pastor1,Pastor2} appears renormalizable only in its extended version that involves infinitely many coupling constants; see~\cite{US,US2,Delamotte,Vestnik,Stat} for discussion.} 

In addition to~(\ref{forceD}), the authors of~\cite{Pastor1,Pastor2} studied the case of the time-independent (spatially quenched) noise with correlation function
\begin{equation}
\langle f_{s}(x)f_{s}(x') \rangle = D_{0}\,
\delta^{(d)}({\bm x}-{\bm x}'), \quad D_0>0.
\label{forceStat}
\end{equation}
It turned out that the model~\cite{Pastor1,Pastor2} predicts nontrivial scaling behavior only for the case of the spatially quenched noise~(\ref{forceStat}). Moreover, in this case the nontrivial behavior is nonuniversal~\cite{Delamotte}.

Originally, the noise~(\ref{forceStat}) was proposed in~\cite{Caldarelli} to reflect the existence of nonerodible (``quenched'') regions of landscape in the problem of erosion. This choice was motivated by the experimental results that had revealed that heterogeneity of the soil is likely the main factor leading to scaling in erosion~\cite{Czi}. The spatially quenched noise~(\ref{forceStat}) and its more general form that depends on the field $h$ [see Eq.~\eqref{eq1}] were also studied in~\cite{Hinri,KimKim,KimKimKim,Nara}. In particular, the connection of this noise to nonuniversality in relation to directed percolation was discussed in~\cite{Janssen,Moreira,Webman}.

In general, the random noise is an essential part of a model. It incorporates various random processes that affect the system while satisfying underlying symmetries of the problem, e.g., the Galilean symmetry. Thus, the choice of the noise is one of the key steps in the model construction. 
Then, it is natural to expect that the change of the noise would greatly affect the
critical behavior of the model. 

However, the exact effect of the type of noise on the scaling behavior is difficult to predict. For example, analysis of the Hwa--Kardar model~(\ref{eq1}) with the spatially quenched noise~(\ref{forceStat}) did not reveal any universality classes with unexpected features~\cite{Vestnik,Stat}. It was shown in~\cite{AGM,AGMK} that the stochastic Navier--Stokes equation with temporally correlated noise reveals the same scaling properties as if the noise was white in-time. 
On the other hand, it was recently reported that temporally correlated noise in the Kardar--Parisi--Zhang model causes anomalous scaling behavior~\cite{KPZan}. 
So this effect seems to be an important avenue to explore. 

The critical behavior of the system can also be greatly affected by the turbulent flows present in the environment; see, \mbox{e.g.,~\cite{Satten,Satten1,Onuki,Onuki2,Beysens,Ruiz,Nelson,AHH,Alexa,AIK,AKM}}. The advection by the velocity field ${\bm v}(x)$ can be introduced by the ``minimal'' replacement in the Eq.~(\ref{eq1}):
\begin{equation}
\partial_{t} \to \nabla_{t} = \partial_{t} + ({\bm v}\cdot \boldsymbol{ \partial}).
\label{nabla}
\end{equation}
Here $\nabla_{t}$ is the Galilean covariant (Lagrangian) derivative.

The Hwa--Kardar model~(\ref{eq1}) with turbulent flow was studied in~\cite{AK1}. 
Since the model~(\ref{eq1}) has an intrinsic strong anisotropy (preferred direction for the sand transport) it is natural to use an anisotropic ensemble for velocity statistics, too. In~\cite{AK1} 
the noise was chosen in the form~(\ref{forceD}) while a $d$-dimensional generalization of the
velocity ensemble with vanishing correlation time, introduced earlier by Avellaneda and Majda~\cite{AM,AM1} was employed as velocity statistics. This ensemble can be considered as 
an anisotropic version of the Kazantsev--Kraichnan ``rapid-change'' ensemble. The latter attracted enormous attention 
on the turn of the Millennium
within the turbulent community because of the deep insight it offered into the origins of intermittency and anomalous scaling in fluid turbulence; see~\cite{FGV} and references therein.

In this paper we try to move ahead by using a more realistic version of the aforementioned Avellaneda--Majda ensemble that incorporates finite correlation time. Let us describe it in detail: following~\cite{AM,AM1} and subsequent works (see also~\cite{AM2} and~\cite{AM3,AM4,AM5,AM6,AM7,AM8,Glimm,Walls}) we take 
the velocity field in the form
\begin{equation}
{\bm v} = {\bm n} \,v(t, {\bm x}_{\bot}),
\label{arrow}
\end{equation}
where $v(t, {\bm x}_{\bot})$ is a scalar function.
Thus defined, the velocity field describes incompressible fluid:
the function $v(t, {\bm x}_{\bot})$ does not depend on $x_{\parallel}$, therefore, $(\boldsymbol{ \partial}\cdot {\bm v}) = \partial_{\parallel} \,v(t, {\bm x}_{\bot}) = 0$.
This velocity ensemble was studied earlier (also in the short-correlated version) in~\cite{Alexa,Alexa2,Lesha2,Lesha3} in connection with the effects of turbulent motion on the dynamic critical behavior and in~\cite{shark,AG1,AG2,AG1a,AG2a} in connection with the problem of anomalous scaling of passively advected scalar and vector fields.
It can also be viewed as an anisotropic modification of the ensemble studied in~\cite{Ant1,Ant2,Ant3,Ant4} in connection with the anomalous scaling in fluid turbulence
and in~\cite{AHH,Alexa,AIK,AKM} in connection with the effects of turbulent motion on the critical behavior.

The amplitude velocity coefficient $v(t, {\bm x}_{\bot})$ has Gaussian distribution with  zero mean and prescribed pair correlation function
\begin{equation}
\langle v(t,{\bm x}_{\bot})\, v(t',{\bm x}_{\bot}') \rangle = \int\frac{d\omega}{2\pi}\int_{k>m}\frac{d{\bm k}}{(2\pi)^{d}}\,e^{i{\bm k}({\bm x}-{\bm x'})-i\omega(t-t')}\,B_v(\omega,{\bm k}),
\label{veloc1}
\end{equation}
where
\begin{equation}
B_{v} (\omega,{\bm k})=2\pi\delta(k_{\|})\, B_0\,\frac{k_{\bot}^{5-d-(\xi+\eta)}}{\omega^2+\left(\alpha_0\nu_{\bot 0}k_{\bot}^{2-\eta}\right)^2}.
\label{veloc2}
\end{equation}
Here $k_{\bot}=|{\bm k}_{\bot}|$, while the infrared
(IR) regularization in the form of the sharp cutoff 
$k_{\bot}>m$ is chosen for convenience.
Other parameters beside $m$ in Eqs.~\eqref{veloc1}~-- \eqref{veloc2} include constant positive amplitude factor $B_{0}$, a new parameter $\alpha_0$ needed for dimensional consistency, and two arbitrary exponents $\xi$ and $\eta$. The role of the exponent $\xi$ could be understood from the asymptotic law of the one-dimensional velocity energy spectrum: 
\begin{equation}
{\cal E} ({\bm k_{\bot}}) \sim k_{\bot}^{d-2}\,B_0\, \int \frac{d\omega}{2\pi}  \frac{k_{\bot}^{5-d-(\xi+\eta)}}{\omega^2+\left(\alpha_0\nu_{\bot 0}k_{\bot}^{2-\eta}\right)^2} =
\frac{B_{0}}{2\alpha_{0}\nu_{\bot 0}}\,k_{\bot}^{1-\xi}.
\label{spektr}
\end{equation}
The exponent $\eta$, on the other hand, appears in the dispersion law
\begin{equation}
\omega({\bm k_{\bot}}) \sim k_{\bot}^{2-\eta}.
\label{dispers}
\end{equation}
The notation $z=2-\eta$ is sometimes used in the literature instead; see, e.g.,~\cite{AM1}.

The specific choice of the velocity correlation function~(\ref{veloc1}), (\ref{veloc2}) can be justified by connection with the stochastic Navier--Stokes equation~\cite{shark}. The substitution~(\ref{arrow}) ``kills'' the nonlinearity in the Navier--Stokes equation:
$({\bm v}\cdot \boldsymbol{\partial})\, v_i = n_i\, v(t, {\bm x}_{\bot} )\, \partial_{\parallel}
v(t, {\bm x}_{\bot} ) =0$.
The equation becomes linear and, thus, determines a Gaussian distribution known as the Ornstein--Uhlenbeck process~\cite{Ito1,Ito2}.
An appropriate power-like choice of the
effective viscosity coefficient and the correlation function of the stirring force leads to the correlation function~(\ref{veloc1}), (\ref{veloc2}). For a more detailed discussion, see Sec.~9 in~\cite{shark}. 

In contrast to the rapid-change model, where correlation function depends on time as $\delta(t-t')$ and do not depend on the frequency $\omega$, our choice~(\ref{veloc1})~-- (\ref{veloc2}) has a power dependence on $\omega$.
This means that it is a colored noise with finite correlation time. 
Depending on $\alpha_0$ and $B_0$ it allows for two special cases interesting on their own.
The limit $\alpha_{0}\to 0$ at fixed $B_0 / \alpha_0$
corresponds to the case of ``frozen'' or ``quenched'' velocity field that does not depend on time. The correlator~(\ref{veloc1}) then
turns into $B_{v} \sim \delta(\omega) k^{3-d-\xi}$.
The limit $\alpha_{0}\to \infty$
at fixed $B_0 / \alpha_0^2$ returns us to the vanishing correlation time (``rapid-change'' case)
where $\langle v(t,{\bm x}_{\bot})\, v(t',{\bm x}_{\bot}') \rangle\sim\delta(t-t') / k_{\bot}^{d-1+{\widetilde\xi}}$ and ${\widetilde\xi}=\xi-\eta$. The exponent $0< {\widetilde\xi} <2 $ is, in a sense, a H\"{o}lder's exponent that indicates ``roughness'' of the velocity field. A smooth velocity is associated with the ``Batchelor limit'' ${\widetilde\xi}\to2$ while the most realistic velocity corresponds to the Kolmogorov values ${\widetilde\xi}=4/3$ and $\eta=4/3$~\cite{FGV}. 
Thus, the Kolmogorov values of the exponents $\xi$ and $\eta$ are $8/3$ and $4/3$, respectively.

In the present paper, we study two models of SOC with field theoretic RG approach. The first model consists of the stochastic equation~(\ref{eq1}) with the white in-time noise~(\ref{forceD}) subjected to the turbulent stirring~(\ref{nabla})~-- (\ref{veloc2}). The second model differs from the first one only in the choice of the random noise in the equation~(\ref{eq1}), i.e., the spatially quenched noise~(\ref{forceStat}) is used instead of the white noise~(\ref{forceD}). As we will see, obtained results are completely different.
Both models can be reformulated as quantum field theories so their possible large-scale, long-distance asymptotic regimes are associated with IR attractive fixed points of the RG equations.
\pagebreak

There are two different ways to organize this paper. On the one hand, we analyze two specific  models and obtain specific results so it is possible to present these two models separately, i.e., in series. On the other hand, we want to stress how the type of the noise affects the results; for this reason it is more convenient to present two models in parallel. We chose the latter way. What is the most interesting is that there are no significant differences between the two models even in the obtained $\beta$ functions: for both models they are very similar to each other, and it is impossible to predict the essential difference in the following analysis at a first glance. In our opinion, this is the most interesting issue from theoretical viewpoints, and, thus, we chose the organisation of the paper that highlighted it. 
We hope that we will not cause too much inconvenience to the reader with the use of repeating symbols for different cases: since the starting equation~\eqref{eq1} and some others are the same for both models, this is unavoidable.

Consequently, the paper is organized as follows.
In Sec.~\ref{sec:QFT1} the field theoretic formulations of the models are presented and Feynman diagrammatic techniques are introduced. 
In Sec.~\ref{sec:Reno1} renormalization of the models (divergent Green functions, renormalized actions and constants $Z$ needed for multiplicative renormalization) is discussed.
Sec.~\ref{sec:Reno2} is devoted to the RG equation, RG functions and IR attractive 
fixed points related to them. 
In Sec.~\ref{sec:DimeNS1} the critical scaling behavior and critical dimensions in different scaling regimes are discussed. 
Sec.~\ref{sec:Conc} is reserved for conclusions. The main result is that the pattern of the fixed points and their regions of stability
for the model with the spatially quenched noise is much more complicated than their
counterparts for the model with the white noise. 

Appendices~\ref{App1} and~\ref{App2} contain some details of the calculations. Since it is a technical point, we do not discuss any details of the calculations in the main text; herewith, it may be useful or interesting at some point to see them.

%%%%%%%%%%%%%%%%%%%%%%%%%%%%%%%%%%%%%%%%%%%%%%%%%%%%%%%%%%%%%%%%%%%%%%%%%%%%%%%%%%%%%%%%%%%%%%%%%%%%%%%%%%%%%%%%%%%%%%%%%%%%%%%%%%%%%%%%%%%%%%%%%%%%%%%%%%%%%%%%%%%%%%%%%%%%%%%%%%%%%%%%%%

\section{Field theoretic formulation of the models}
\label{sec:QFT1}

%%%%%%%%%%%%%%%%%%%%%%%%%%%%%%%%%%%%%%%%%%%%%%%%%%%%%%%%%%%%%%%%%%%%%%%%%%%%%%%%%%%%%%%%%%%%%%%%%%%%%%%%%%%%%%%%%%%%%%%%%%%%%%%%%%%%%%%%%%%%%%%%%%%%%%%%%%%%%%%%%%%%%%%%%%%%%%%%%%%%%%%%%%

From now on, every section is organised as follows: we start with the model that involves the white noise~(\ref{forceD}) which we refer to as Model~1 in the text. Then we consider the model with the spatially quenched noise~\eqref{forceStat}; this model is referred to as Model~2.

According to the general theorem, any stochastic differential equation of the type
(\ref{eq1})~-- (\ref{forceStat}), first-order in the time derivative,
is equivalent to a field theoretic model with certain action 
functional~${\cal S}(\Phi)$; see, e.g., 
\cite{MSR,MSR1,MSR2,MSR11,MSR111,MSR12,MSR121} and the monographs~\cite{Zinn,Book3}.\footnote{It is essential here that the 
interaction term depends only on the fields and their spatial derivatives of arbitrary order at a single moment  $t$.}
This equivalence means that statistical averages of random quantities in the initial stochastic problem coincide with functional averages with the weight 
$\exp {\cal S}(\Phi)$.\footnote{In fact, the main idea of this formalism dates back to the seminal works of Onsager and Machlup on irreversible stochastic processes~\cite{OM,OM1}.}
This idea appears to be very fruitful and allows one to apply the well-known techniques of quantum field theory, like Feynman diagrammatic techniques, renormalization and RG equation, operator product expansion, etc., to problems of statistical physics. 

%%%%%%%%%%%%%%%%%%%%%%%%%%%%%%%%%%%%%%%%%%%%%%%%%%%%%%%%%%%%%%%%%%%%%%%%%%%%%%%%%%%%%%%%%%%%%%%%%%%%%%%%%

\subsection{Model 1: The model with the white noise}
\label{App1Z2}

%%%%%%%%%%%%%%%%%%%%%%%%%%%%%%%%%%%%%%%%%%%%%%%%%%%%%%%%%%%%%%%%%%%%%%%%%%%%%%%%%%%%%%%%%%%%%%%%%%%%%%%%%

The action functional ${\cal S}(\Phi)$ mentioned above  for the stochastic problem~(\ref{eq1}), (\ref{forceD}), \eqref{nabla}, (\ref{veloc1}) involves the extended set of fields 
$\Phi = \{ h', h, {\bm v} \}$ and reads
\begin{equation}
{\cal S}(\Phi)=\frac{1}{2}\,h' D_0\, h'+h'\left\{-\partial_{t}h-v\,\partial_{\parallel}h+
\nu_{\bot 0}\, \boldsymbol{\partial}_{\bot}^{2} h +
\nu_{\parallel 0}\, \partial_{\parallel}^{2} h -\partial_{\parallel} h^{2}/2\right\}+ {\cal S}_v.
\label{action}
\end{equation}
Here $h'$ is the auxiliary (response) field and all the integrations over 
$x=\{t,{\bm x}\}$ and
summations over the vector indices are implied;
for instance,
\begin{equation}
\frac{1}{2}h'D_0 \,h' = \frac{1}{2}\int dt\, d {\bm x}\,
h'(t,{\bm x})\,h'(t,{\bm x}).
\label{actionX}
\end{equation}
The term ${\cal S}_v$ describes the Gaussian averaging over the velocity field ${\bm v}$:
\begin{equation}
{\cal S}_v= \frac{1}{2}\,
\int dt \,d{\bm x}_{\bot} d{\bm x}_{\bot}'\, v(t,{\bm x}_{\bot})\,
{\widetilde B}^{-1}_{v} ({\bm x_{\bot}}-{\bm x'_{\bot}}) \,v(t,{\bm x}_{\bot}'),
\label{Sv}
\end{equation}
where ${\widetilde B}^{-1}_{v}$ is the kernel of the linear operation $B^{-1}_{v}$ which is the inverse operation for the $B_{v}$ in~(\ref{veloc2}).

Feynman diagrammatic technique for the theory~(\ref{action}) involves four bare propagators. The velocity propagator $\langle v v \rangle_{0}$ is defined in~(\ref{veloc1}).
Other four propagators that contain the height field $h$ and response field $h'$ in the frequency-momentum representation read
\begin{eqnarray}
\langle hh \rangle_{0}=\frac{D_{0}} { \omega^{2} + \epsilon^{2}(k) }, \quad
\langle hh' \rangle_{0} &=& \langle h'h
\rangle_{0}^{*}
= \frac{1}{-{i} \omega+ \epsilon(k)}, \quad
\langle h'h'\rangle_{0}=0,
\label{lines3}
\end{eqnarray}
where we denote $\epsilon(k)=\nu_{\parallel 0}\, k_{\parallel}^2+\nu_{\bot 0}\, k_{\bot}^2$.
\pagebreak

The nonlinear terms $-h'\partial_{\parallel} h^{2}/2$ and $-h'(v\partial_{\parallel})h$ define the vertices $V_{h'hh}$ and $V_{h'vh}$.
It is convenient to define the corresponding
coupling constants $g_{0}$ and $w_{0}$ by the relations
\begin{equation}
D_0=g_0\,\nu_{\| 0}^{3/2}\,\nu_{\bot 0}^{(d_L-1)/2}, \quad B_0=w_0\,\nu_{\| 0}\,\nu_{\bot 0}^2,
\label{D0}
\end{equation}
where $d_L$ is logarithmic dimension of the model. Then, canonical dimension analysis (see Sec.~\ref{sec:Reno1} for details) gives $g_{0} \sim \ell^{-\varepsilon}$ and
$w_{0} \sim \ell^{-\xi-\eta}$,
where $\ell$ sets the smallest length scale in the problem (ultraviolet cutoff) and $\varepsilon= 4-d$.
The parameter $\alpha_0 \sim \ell^{-\eta}$
should be considered alongside the coupling constants. Indeed, although it is not an expansion parameter in the perturbation theory, the RG function will depend on its renormalized analog.

%%%%%%%%%%%%%%%%%%%%%%%%%%%%%%%%%%%%%%%%%%%%%%%%%%%%%%%%%%%%%%%%%%%%%%%%%%%%%%%%%%%%%%%%%%%%%%%%%%%%%%%%%

\subsection{Model 2: the model with the spatially quenched noise}
\label{App2Z2}

%%%%%%%%%%%%%%%%%%%%%%%%%%%%%%%%%%%%%%%%%%%%%%%%%%%%%%%%%%%%%%%%%%%%%%%%%%%%%%%%%%%%%%%%%%%%%%%%%%%%%%%%%

Now let us turn to Model~2. As the previous one, it can be reformulated as a field theory
of the set of three fields $\Phi = \{ h', h, {\bm v} \}$. The action functional has the same form as Eq.~\eqref{action} with the only difference:
the first term now reads
\begin{equation}
\frac{1}{2}h' D_0\,h'=\frac{1}{2}\int dt\, dt'\int d{\bm x}\,h'(t',x)\,D_0\,h'(t,x)
\label{S1}
\end{equation}
with the double integration over the time variables.

As Model~1, Model~2 also involves five bare propagators. The propagator $\langle vv \rangle_{0}$ is still defined in~(\ref{veloc1}); another four propagators in the frequency-momentum representation read
\begin{eqnarray}
\langle hh \rangle_{0}=
\frac{2\pi \delta(\omega)\,D_{0}}{\epsilon^{2}(k)}, \quad 
\langle hh' \rangle_{0} = \langle h'h
\rangle_{0}^{*}
= \frac{1}{-{i} \omega+ \epsilon(k)
}, \quad
\langle h'h'\rangle_{0}=0,
\label{stlines3}
\end{eqnarray}
where $\epsilon(k)$ is defined by linear part of Hwa--Kardar equation and, therefore, is the same as for Model~1; see Eq.~\eqref{lines3}.

As before, the theory involves
two vertices related to the interaction terms
and three coupling constants: $g_{0}$ and $w_{0}$ defined by~\eqref{D0}
and $\alpha_0$.
From canonical dimension analysis (see Sec.~\ref{sec:Reno1}) it follows that $g_{0} \sim \ell^{-\tilde\varepsilon}$,
$w_{0} \sim \ell^{-\xi-\eta}$, and $\alpha_{0} \sim \ell^{-\eta}$ with $\tilde\varepsilon=6-d$.

%%%%%%%%%%%%%%%%%%%%%%%%%%%%%%%%%%%%%%%%%%%%%%%%%%%%%%%%%%%%%%%%%%%%%%%%%%%%%%%%%%%%%%%%%%%%%%%%%%%%%%%%%%%%%%%%%%%%%%%%%%%%%%%%%%%%%%%%%%%%%%%%%%%%%%%%%%%%%%%%%%%%%%%%%%%%%%%%%%%%%%%%%%

\section{Renormalization of the models}
\label{sec:Reno1}

%%%%%%%%%%%%%%%%%%%%%%%%%%%%%%%%%%%%%%%%%%%%%%%%%%%%%%%%%%%%%%%%%%%%%%%%%%%%%%%%%%%%%%%%%%%%%%%%%%%%%%%%%%%%%%%%%%%%%%%%%%%%%%%%%%%%%%%%%%%%%%%%%%%%%%%%%%%%%%%%%%%%%%%%%%%%%%%%%%%%%%%%%%

%%%%%%%%%%%%%%%%%%%%%%%%%%%%%%%%%%%%%%%%%%%%%%%%%%%%%%%%%%%%%%%%%%%%%%%%%%%%%%%%%%%%%%%%%%%%%%%%%%%%%%%%%

%\subsection{General remarks}

%%%%%%%%%%%%%%%%%%%%%%%%%%%%%%%%%%%%%%%%%%%%%%%%%%%%%%%%%%%%%%%%%%%%%%%%%%%%%%%%%%%%%%%%%%%%%%%%%%%%%%%%%

Ultraviolet (UV) divergences are determined through canonical dimensions analysis (``power counting''), see, e.g.,~\cite{Amit,Zinn,Book3}. Let us briefly detail the process. Firstly, one needs to find canonical dimensions of the fields and parameters of the theory. The strongly anisotropic dynamic theories like Model~1 and Model~2 have three independent scales: the time scale $T$ and two length scales (in the corresponding subspaces) $L_{\bot}$ and $L_{\parallel}$. Thus, a quantity $F$ is described by three canonical dimensions:
\[ [F] \sim [T]^{-d_{F}^{\omega}} [L_{\bot}]^{-d_{F}^{\bot}}
[L_{\parallel}]^{-d_{F}^{\parallel}}. \]
The total canonical dimension $d_{F}$ is a sum of the doubled frequency dimension $d_{F}^{\omega}$ and the momentum dimensions $d_{F}^{\bot}$ and $d_{F}^{\parallel}$: $d_{F}=
d_{F}^{\bot} + d_{F}^{\parallel} +2\,d_{F}^{\omega}$. The free theory relation $\partial_{t}\propto\partial^{2}_{\bot} \propto \partial^{2}_{\parallel}$ explains the factor $2$.

As each term of the action~(\ref{action}) is completely dimensionless ($\sim [T]^{0} [L_{\bot}]^{0}
[L_{\parallel}]^{0}$), the canonical dimensions can be easily calculated; the normalization conditions $d_{{k_{\bot}}}^{\bot}= -d_{x_{\bot}}^{\bot}=1$,
$d_{{k_{\bot}}}^{\parallel}=-d_{x_{\bot}}^{\parallel}=0$,
$d_{{k_{\bot}}}^{\omega} = d_{k_{\parallel}}^{\omega}=0$,
$d_{\omega }^{\omega }=-d_t^{\omega }=1$ are assumed.

%%%%%%%%%%%%%%%%%%%%%%%%%%%%%%%%%%%%%%%%%%%%%%%%%%%%%%%%%%%%%%%%%%%%%%%%%%%%%%%%%%%%%%%%%%%%%%%%%%%%%%%%%

\subsection{Model 1: the model with the white noise}
\label{App1Z}

%%%%%%%%%%%%%%%%%%%%%%%%%%%%%%%%%%%%%%%%%%%%%%%%%%%%%%%%%%%%%%%%%%%%%%%%%%%%%%%%%%%%%%%%%%%%%%%%%%%%%%%%%

The canonical dimensions for Model~1 are presented in Table~\ref{t1}. The parameter $\mu$ is the renormalization mass, i.e., the reference momentum scale defined by its canonical dimensions~\cite{Book3}.

\begin{table*}[t]
\caption{Canonical dimensions of the fields and the parameters in Model~1; $\varepsilon= 4-d$.}
\label{t1}
\begin{ruledtabular}
\begin{tabular}{c||c|c|c|c|c|c|c|c|c|c|c}
$F$&$h'$&$h$&$D_0$&$\nu_{\| 0}$&$\nu_{\bot 0}$&$v$&$B_0$&$\alpha_0$&$g_0$&$w_0$&$\mu$, $m$\\
\hline
$d^{\omega}_F$&$-1$&$1$&$3$&$1$&$1$&$1$&$3$&$0$&$0$&$0$&$0$\\
%\hline
$d^{\|}_F$&$2$&$-1$&$-3$&$-2$&$0$&$-1$&$-2$&$0$&$0$&$0$&$0$\\
%\hline
$d^{\bot}_F$&$d-1$&$0$&$1-d$&$0$&$-2$&$0$&$\xi+\eta-4$&$\eta$&$\varepsilon$&$\xi+\eta$&$1$\\
%\hline
$d_F$&$d-1$&$1$&$4-d$&$0$&$0$&$1$&$\xi+\eta$&$\eta$&$\varepsilon$&$\xi+\eta$&$1$\\
\end{tabular}
\end{ruledtabular}
\end{table*}

From Table~\ref{t1} it follows that the model is logarithmic (all the coupling constants are dimensionless, or, in other words, all the interactions are marginal in the sense of Wilson)
at $\varepsilon=\xi=\eta=0$, where $\varepsilon=4-d$.
Thus, these three exponents will serve as the expansion parameters in the RG theory.

Once canonical dimensions are found, the UV divergences can be analyzed. The UV divergence index of an arbitrary 1-irreducible Green function
$\Gamma = \langle\Phi \cdots \Phi \rangle _{1-ir}$ is given by the
expression
\begin{equation}
\delta_{\Gamma }= d+2- N_{h'}\, d_{h'} - N_{h}\, d_{h} - N_{{v}}\, d_{{v}}\,|_{\varepsilon=\xi=\eta=0},
\label{dGamma}
\end{equation}
where $N_{h},\,N_{h'},\,N_{\bm v}$ are the numbers of the corresponding fields in the function $\Gamma$.
\pagebreak

If $\delta_{\Gamma}$ is a nonnegative integer, then the function $\Gamma$ may contain superficial UV divergences. Table~\ref{t1} and expression~(\ref{dGamma}) gives
\begin{equation}
\delta_{\Gamma}= 6 - 3 N_{h'} - N_{h} -N_{v}.
\label{IndeX}
\end{equation}

There are additional
considerations that should be taken into account when analyzing UV divergences. Firstly, since both vertices $V_{h'hh}$ and $V_{h'vh}$ allow to move derivative $\partial_\parallel$ onto the field $h'$ the real index of divergence reads
\begin{equation}
\delta'_{\Gamma}=  \delta_{\Gamma} - N_{h'}.
\end{equation}
Moreover, all 1-irreducible Green functions without response field $h'$  involve closed circuits of retarded propagators $\langle h'h \rangle_{0}$ and, thus, vanish~\cite{Book3}. So, $N_{h'}\geq 1$.

The Galilean symmetry usually forbids some of the counterterms allowed by power counting and, therefore, reduces the number of counterterms.
However, the correlation function~(\ref{veloc1}) does not contain the Dirac function $\delta(t-t')$ necessary for Galilean symmetry. This lack of symmetry may result in some ``interesting physical pathologies''~\cite{synth}. In the present case, though, due to the strong anisotropy of the theory~(\ref{action}) and incompressibility of the velocity, the action~(\ref{action}) is invariant under the following Galilean transformations
\begin{equation}
h(t,{\bm x}) \to h(t,{\bm x}+{\bm u}\,t), \quad
h'(t,{\bm x}) \to h'(t,{\bm x}+{\bm u}\,t), \quad
{\bm v}(t, {\bm x}) \to {\bm v}(t, {\bm x} +{\bm u}\,t) - {\bm u},
\label{GT}
\end{equation}
where ${\bm u}={\bm n}\, u$, which can be verified by the direct substitution. 
Expression~\eqref{GT} means that the scalar velocity changes as $v(t, {\bm x}_{\bot}) \to v(t, {\bm x}_{\bot})-u$
and ${\bm x}_{\bot}$ remains unchanged in all of the fields in~(\ref{GT}).
This symmetry can be viewed as a residue of the full-scale Galilean symmetry that survived the substitution~(\ref{arrow}) made in the Navier--Stokes equation.\footnote{We stress that in the isotropic version, the Gaussian velocity ensemble with a finite correlation time is not Galilean covariant; for a discussion see, e.g.,~\cite{synth,Ant1,Ant2,shark}.}

In our case this observation forbids counterterms for 1-irreducible functions with the field $v$, namely $\langle h'v \rangle_{1-ir} $ with $\delta_\Gamma=2$, $\langle h'hv \rangle_{1-ir} $ with $\delta_\Gamma=1$, and $\langle h'vv \rangle_{1-ir} $ with $\delta_\Gamma=1$. 
Moreover, there are two types of graphs for function $\langle h'hh \rangle_{1-ir} $ with $\delta_\Gamma=1$: the one with propagator $\langle vv \rangle $ inside the core (integrand) and the one without. The former is trivially equal to zero while the core of the graphs of the latter fully coincides with similar cores for the function $\langle h'hv \rangle_{1-ir} $. This means that the Galilean symmetry, in fact, forbids the possible counterterm for function $\langle h'hh \rangle_{1-ir}$, too. 

Taking all of the above into account, we can ascertain that only one counterterm has to be considered which is $h'\partial_{\parallel}^2 h$ that appears
from the 1-irreducible function $\langle h'h \rangle_{1-ir} $ with $\delta_\Gamma=2$.
This means that Model~1 is renormalizable and renormalized action reads
\begin{equation}
{\cal S}_R ( h,h',v)=\frac{1}{2}h' D \,h'+h'\left\{-\,\partial_{t}h
-\,v\,\partial_{\parallel}h
+ \,\nu_{\bot} \boldsymbol{\partial}_{\bot}^{2} h + Z_{\nu_{\parallel}}\,\nu_{\parallel}\,
\partial_{\parallel}^{2} h -
\,\partial_{\parallel} h^{2}/2\right\}+\,{\cal S}_v.
\label{ren}
\end{equation}
This renormalization can be reproduced by multiplicative renormalization of the parameters
\begin{equation}
g_0 = \mu^{\varepsilon}\, g\, Z_g, \quad
w_0 = \mu^{\zeta+\eta}\, w\, Z_w, \quad
\alpha_0 = \alpha \, \mu^{\eta}, \quad
\nu_{\parallel\, 0} = \nu_{\parallel} \, Z_{\nu_{\parallel}}, \quad
\nu_{\bot\,0} = \nu_{\bot}.
\label{Multi}
\end{equation}
Here $g$, $w$, etc., are renormalized counterparts of the bare parameters $g_0$, $w_0$, etc.; $\mu$ is renormalization mass, an additional parameter of the renormalized theory (see, e.g.,~\cite{Zinn,Book3}). Due to the fact that there is only one counterterm, the fields $h, h'$, and ${\bm v}$ are not renormalized and following relations hold true:
\[ Z_g = Z_{\nu_{\parallel}}^{-3/2}, \quad Z_w =
Z_{\nu_{\parallel}}^{-1}. \]

The renormalization constant $Z_{\nu_{\parallel}}$ can be calculated in the double series in $g$ and $w$.
In the minimal subtraction~(MS) scheme all the renormalization constants have the forms ``$Z=1+$ only poles in $\varepsilon$, $\xi$ and their combinations.''
The leading-order (one-loop) calculation gives
\begin{equation}
Z_{\nu_{\parallel}}=1-\frac{1}{2\alpha\,(1+\alpha)}\,\frac{w}{\xi}\,-\,\frac{3}{16}\,\frac{g}{\varepsilon}
\label{ZM1}
\end{equation}
with the corrections of higher orders in $g$ and $w$. 
Here and below we redefined the coupling constant $g\to gS_d/(2\pi)^d$ where $S_d=2\pi^d/\Gamma(d/2)$ is the area of the unit sphere in the $d$-dimensional space; the same redefinition is also true for the second coupling constant $w$. Details of the calculations can be found in Appendix~\ref{App1}.

%%%%%%%%%%%%%%%%%%%%%%%%%%%%%%%%%%%%%%%%%%%%%%%%%%%%%%%%%%%%%%%%%%%%%%%%%%%%%%%%%%%%%%%%%%%%%%%%%%%%%%%%%

\subsection{Model 2: the model with the spatially quenched noise}
\label{App2Z}

%%%%%%%%%%%%%%%%%%%%%%%%%%%%%%%%%%%%%%%%%%%%%%%%%%%%%%%%%%%%%%%%%%%%%%%%%%%%%%%%%%%%%%%%%%%%%%%%%%%%%%%%%

Now let us turn again to Model~2.
Canonical dimensions for Model~2 are presented in Table~\ref{t2}. The only difference between the two sets of canonical dimensions is the dimension of the parameter $D_0$ which leads to a different dimension of the coupling constant $g_0$.  
This in turn leads to the shift of the logarithmic dimension of the model: now all of the couplings are dimensionless at $\tilde\varepsilon=\xi=\eta=0$ where $\tilde\varepsilon=6-d$. 

\begin{table*}[t]
\caption{Canonical dimensions of the fields and the parameters in Model~2; $\tilde\varepsilon= 6-d$.}
\label{t2}
\begin{ruledtabular}
\begin{tabular}{c||c|c|c|c|c|c|c|c|c|c|c}
$F$&$h'$&$h$&$D_0$&$\nu_{\| 0}$&$\nu_{\bot 0}$&$v$&$B_0$&$\alpha$&$g_0$&$w_0$&$\mu$, $m$\\
\hline
$d^{\omega}_F$&$-1$&$1$&$4$&$1$&$1$&$1$&$3$&$0$&$0$&$0$&{$0$}\\
$d^{\|}_F$&$2$&$-1$&$-3$&$-2$&$0$&$-1$&$-2$&$0$&$0$&$0$&{$0$}\\
$d^{\bot}_F$&$d-1$&$0$&$1-d$&$0$&$-2$&$0$&$\xi+\eta-4$&$\eta$&$\tilde\varepsilon$&$\xi+\eta$&{$1$}\\
$d_F$&$d-1$&$1$&$6-d$&$0$&$0$&$1$&$\xi+\eta$&$\eta$&$\tilde\varepsilon$&$\xi+\eta$&{$1$}\\
\end{tabular}
\end{ruledtabular}
\end{table*}

The UV divergence index of an arbitrary 1-irreducible Green function $\Gamma$ is given by Eq.~(\ref{dGamma}) and reads
\begin{equation}
\delta_{\Gamma}= 8 - 5 N_{h'} - N_{h} -N_{v}.
\label{stIndeX}
\end{equation}

The spatially quenched noise~\eqref{forceStat} destroys Galilean symmetry~\eqref{GT} which is true for Model~1. This fact can be checked directly: the term~\eqref{S1} is not invariant under the transformations~\eqref{GT}. 
Thus, in contrast to Model~1, we have to deal with all five types of divergent functions:  $\langle h'h \rangle_{1-ir} $, $\langle h'v \rangle_{1-ir} $, $\langle h'hv \rangle_{1-ir} $, $\langle h'hh \rangle_{1-ir} $, and $\langle h'vv \rangle_{1-ir} $.

In the same time, since the propagator $\langle hh \rangle_{0}$ in Eq.~\eqref{stlines3} is proportional to $\delta(\omega)$, Model~2 has an additional feature that leads to a reduction of counterterms:
when a diagram involves $n\geq2$ inner lines $\langle hh \rangle_{0}$, it with necessity has $(n-1)$ delta functions of external frequencies $\delta(\Omega_{i})$ as factors. Each factor contributes $d_{\delta(\Omega_{i})}=-2$ to the divergence index while being unrelated to the momenta divergence. Thus, the real index of divergence has an additional term $2(n-1)$ and reads
\begin{equation}
\delta_{\Gamma}''=\delta_{\Gamma}-N_{h'}+2(n-1).
\label{stIndeXR}
\end{equation}
The possible ``dangerous'' function of such type is $\langle h'h' \rangle_{1-ir}$. It has formal index of divergence $\delta_\Gamma=-2$ but the one-loop approximation contains the graph with two lines $\langle hh \rangle_{0}$. 
This allows for a possibility that integral over momenta has a logarithmic divergence. However, the situation is safe due to the two vertices $V_{h'hh}$ which are responsible for the term~$-N_{h'}$ in Eq.~\eqref{stIndeXR}. Thus, the real index of divergence for this function $\delta_{\Gamma}''=-2$ and we have no problems with it. 

One more nontrivial observation for this model is worth mentioning. Usually when we state that a Green function is divergent we actually mean that there are divergences of the integrals over momenta, i.e., divergences of the Feynman graphs itself. But integrals over momenta are just a core of the Green functions: they should be contracted with external projectors, propagators or fields.
If transverse vector fields are involved, such a contraction  may lead to an unexpected vanishing of the result. 

Let us consider the function $\langle h'v \rangle_{0}$ whose index of divergence is  $\delta_{\Gamma}''=1$, so, according to the dimensional analysis, we should account for it in the renormalization procedure. 
However, owing to the vertex factor $V_{h'hh}$, each graph for this function is proportional to an external momenta $p_\parallel$. 
This feature along with the property $\partial_{\parallel} \,v(t, {\bm x}_{\bot}) = 0$ [see Eq.~\eqref{arrow}] leads to the fact that $\langle h'v \rangle_{0}=0$ after the contraction of the core of the graph with the external ``tails'' $h'$ and ${\bm v}$. 

The same observation also holds for the function $\langle h'vv \rangle_{0}$: each graph contains two external momenta and the Green function itself involves two vector fields ${\bm v}$. Thus, this function also vanishes, along with the corresponding counterterm. 

The similar observation is no longer true for the function $\langle h'hv \rangle_{0}$: there is still two external momenta but the Green function itself involves only one vector field ${\bm v}$. This means that some nontrivial divergent part survives the contraction. 

The functions that contain four or more fields have negative real index of divergence $\delta'_\Gamma$ and, therefore, are not needed for renormalization procedure from general requirements.

Taking all of the above into account, we can ascertain that three counterterms has to be considered, which are $h'\partial_{\parallel}^2 h$, $h'v\,\partial_{\|}\,h$, and $h'\partial_{\|}h^2$ that appear
from the 1-irreducible functions $\langle h'h \rangle_{1-ir}$, $\langle h'hv \rangle_{1-ir}$, and $\langle h'hh \rangle_{1-ir}$ correspondingly.
Thus, Model~2 is multiplicatively renormalizable and renormalized action reads
\begin{equation}
{\cal S}( h,h',v)=\frac{1}{2}\, h'D\, h'+h'\{-\partial_{t}h
-Z_v\,v\,\partial_{\parallel}h
+ \nu_{\bot} \boldsymbol{\partial}_{\bot}^{2} h + Z_{\nu_{\parallel}}\,\nu_{\parallel}\,
\partial_{\parallel}^{2} h -
Z_{h}\,\partial_{\parallel} h^{2}/2\}+ \,{\cal S}_v.
\label{stren}
\end{equation}
This procedure can be reproduced by multiplicative renormalization of the fields $h\to h\, Z_h$, $h'\to h'\, Z_{h'}$, $v\to v\, Z_v$ and the parameters
\begin{eqnarray}
g_0 = \mu^{\tilde\varepsilon}\, g\, Z_g, \quad
w_0 = \mu^{\zeta+\eta}\, w\, Z_w, \quad
\alpha_0 = \alpha \mu^{\eta}, \quad
\nu_{\parallel\, 0} = \nu_{\parallel} \, Z_{\nu_{\parallel}},
\label{Multi2}
\end{eqnarray}
where $g$, $w$, etc. are renormalized counterparts of the bare parameters and $\mu$ is the renormalization mass. The viscosity $\nu_{\bot\,0}$ remains the same: $\nu_{\bot\,0} = \nu_{\bot}$.
The relations
\begin{equation}
Z_{h'}=Z_h^{-1}, \quad Z_g = Z_h^2\, Z_{\nu_{\parallel}}^{-3/2},
\quad Z_w = Z_v^2\, Z_{\nu_{\parallel}}^{-1} 
\label{ZS1}
\end{equation}
result from the absence of renormalization of the other terms in~(\ref{stren}).

Three independent constants $Z_{\nu_{\parallel}}$, $Z_v$, and $Z_h$ can be calculated in the double series in $g$ and $w$. In the leading order (one-loop approximation) and MS scheme they read
\begin{eqnarray}
Z_v=Z_h=1+\frac{1}{6}\,\frac{g}{\widetilde\varepsilon}; \quad Z_{\nu_{\parallel}}=1-\frac{1}{2\alpha\,(1+\alpha)}\frac{w}{\xi}-\frac{2}{3}\frac{g}
{\tilde\varepsilon}.
\label{ZS2}
\end{eqnarray}
Details of the calculations can be found in Appendix~\ref{App2}.

%%%%%%%%%%%%%%%%%%%%%%%%%%%%%%%%%%%%%%%%%%%%%%%%%%%%%%%%%%%%%%%%%%%%%%%%%%%%%%%%%%%%%%%%%%%%%%%%%%%%%%%%%%%%%%%%%%%%%%%%%%%%%%%%%%%%%%%%%%%%%%%%%%%%%%%%%%%%%%%%%%%%%%%%%%%%%%%%%%%%%%%%%%

\section{Renormalization group, fixed points, and scaling regimes}
\label{sec:Reno2}

%%%%%%%%%%%%%%%%%%%%%%%%%%%%%%%%%%%%%%%%%%%%%%%%%%%%%%%%%%%%%%%%%%%%%%%%%%%%%%%%%%%%%%%%%%%%%%%%%%%%%%%%%%%%%%%%%%%%%%%%%%%%%%%%%%%%%%%%%%%%%%%%%%%%%%%%%%%%%%%%%%%%%%%%%%%%%%%%%%%%%%%%%%

The relation between the initial action functional and the renormalized one $S(\Phi,e_{0})= S_{R}(Z_\Phi\Phi,e,\mu)$, 
where $e$ is the complete set of parameters, yields the fundamental RG differential equation whose coefficients are so-called $\beta$ and $\gamma$ functions (also referred to as RG functions). 
They are defined as
\begin{equation}
\beta_q=\widetilde{\cal D}_{\mu}q, \quad 
\gamma_{F}= \widetilde{\cal D}_{\mu}\ln Z_{F},
\label{RGDef}
\end{equation}
where $F$ denotes any quantity (a field or a parameter) with nontrivial renormalization constant $Z_{F}$ and 
$q=\{g,w,\alpha\}$ is any of the coupling constants.
Here and below ${\cal D}_{x}=x\partial_{x}$ for any variable $x$ and
$\widetilde{\cal D}_{\mu}=\mu\partial_{\mu}$ at fixed bare parameters 
${\nu_{\parallel 0}, \nu_{\bot 0}, w_{0}, g_{0},\alpha_0}$.

The analysis of the RG equations shows that  
the long-time, large-scale asymptotic behavior of a 
given model is governed by the IR attractive fixed points
$q^*$. In our case, the coordinates of the fixed points 
$q^*=\left\{g^{*}, w^{*}, \alpha^{*}\right\}$
are found from the equations  
\begin{equation}
\beta_{g} (g^{*},w^{*},\alpha^{*}) = 0, \quad \beta_{w} (g^{*},w^{*}\alpha^{*})=0, \quad \beta_{\alpha} (g^{*},w^{*},\alpha^{*}) = 0.
\label{points}
\end{equation}
The point is IR attractive (or IR stable) if the real parts of all the eigenvalues $\lambda_{i}$ of the matrix 
\begin{equation}
\Omega_{ij}=\frac{\partial\beta_{i}}{\partial g_{j}}\,\bigg|_{g^{*}, w^{*}, \alpha^{*}}
\label{Omega}
\end{equation}
are positive. This follows from the analysis of the asymptotic behavior of the system of ordinary differential equations for the invariant (``running'') coupling constants in the vicinity of a given fixed point: 
\begin{equation}
  {\cal D}_{s} \overline{q}_i = \beta_i(\overline{q}_j),
  \label{eq:invariant_chrg}
\end{equation}
whose solution as $s=k/\mu\to0$ (IR limit) reads
\begin{equation}
  \overline{q}_i(s,q) \cong q_i^*+ \sum_i c_i\, s^{\lambda_i}.
  \label{Asym}
\end{equation}
Here $c_i$ are some constants, $\lambda_i$ are the eigenvalues of the matrix~\eqref{Omega} and $q=\{q_i\}$ is the set of the coupling constants.

Alternatively, the first-order system of differential equation (a dynamical system)~\eqref{eq:invariant_chrg} gives rises to the possibility of numerical simulation of the RG flow. Such simulation allows one to check the obtained analytical results. Below we present the results of such simulations for the  Model~2: since it has most interesting and entangled pattern of the RG flows, it is very desirable to compare the results of the analytical analysis with the outcome of numerical simulations.

%%%%%%%%%%%%%%%%%%%%%%%%%%%%%%%%%%%%%%%%%%%%%%%%%%%%%%%%%%%%%%%%%%%%%%%%%%%%%%%%%%%%%%%%%%%%%%%%%%%%%%%%%

\subsection{Model 1: the model with the white noise}
\label{sec:eta}

%%%%%%%%%%%%%%%%%%%%%%%%%%%%%%%%%%%%%%%%%%%%%%%%%%%%%%%%%%%%%%%%%%%%%%%%%%%%%%%%%%%%%%%%%%%%%%%%%%%%%%%%%

As there is only one nontrivial independent renormalization constant in Model~1, all the $\beta$ functions  can be expressed through the only anomalous dimension $\gamma_{\nu_{\|}}$:
\begin{eqnarray}
\beta_w=-w\,\left(\xi+\eta-\gamma_{\nu_{\|}}\right), \quad \beta_g=-g\,\left(\varepsilon- \frac{3}{2}\,\gamma_{\nu_{\|}}\right), \quad
\beta_{\alpha}=-\alpha\,\eta.
\label{betagw}
\end{eqnarray}
From Eq.~\eqref{ZM1} it follows that 
\begin{equation}
\gamma_{\nu_{\|}}=\frac{w}{2\alpha\,(1+\alpha)}+\frac{3g}{16}
\label{betagw2}
\end{equation}
with corrections of higher orders in $g$ and $w$ implied. This leads to a following system of $\beta$ functions: 
\begin{eqnarray} \nonumber
\beta_w&=&w\left[-\xi-\eta+\frac{w}{2\alpha(1+\alpha)}+\frac{3g}{16}\right];\\ \nonumber
\beta_g&=&g\left[-\varepsilon+\frac{3w}{4\alpha(1+\alpha)}+\frac{9g}{32}\right];\\ 
\beta_{\alpha}&=&-\alpha\eta. 
\label{BM1}
\end{eqnarray}
It should be noted that the functions $\beta_{g}$ and $\beta_{w}$ in Eq.~(\ref{betagw}) satisfy the exact relation 
\begin{equation}
w\,\beta_{g}-3g\,\beta_{w}/2 =g\,w\, [-\varepsilon+3(\xi+\eta)/2]
\label{relation}
\end{equation}
as a consequence of the fact that both of them involve the same anomalous dimension $\gamma_{\nu_{\|}}$. Thus, the equations~(\ref{points}) are not satisfied for arbitrary values of the exponents $\varepsilon$, $\xi$, and $\eta$ unless one of the coordinates $g^{*}$ or $w^{*}$ is equal to zero (cf.~\cite{Ant1} for the isotropic case).

The analysis of the system~\eqref{BM1} reveals two groups of the fixed points arranged according to the value of $\alpha^*$. The two possible values of $\alpha^*$ are $\alpha^*=0$ and $1/\alpha^*=0$. The first case describes a frozen (or ``quenched'') velocity field, while the second corresponds to the ``rapid-change ensemble''  with vanishing correlation time; see the comments below Eqs.~(\ref{veloc2})~-- (\ref{dispers}). 

Let us start with the former: $\alpha^*=0$, $\lambda_{\alpha}=-\eta$.
It is then convenient to replace the coupling constant $w$ with the new one $w'=w/\alpha$ with the corresponding $\beta$ function \begin{equation}
\beta_{w'}=\frac{\beta_w}{\alpha}-w\,\frac{\beta_{\alpha}}{\alpha^2}=w'\left(-\xi+\frac{w'}{2}+\frac{3g}{16}\right)
\label{BM1a}
\end{equation}
which remains finite at $\alpha\to0$.

The relation~\eqref{relation}, therefore, becomes
\begin{equation}
w'\,\beta_{g}-3g\,\beta_{w'}/2 =g\,w'\, \left(-\varepsilon+3\xi/2\right),
\label{relation1}
\end{equation}
and the system~\eqref{BM1} with the replacement~\eqref{BM1a} allows three possible solutions:

The point (1a) with the coordinates $w'^*=0$, $g^*=0$ and the eigenvalues of the matrix~\eqref{Omega} $\lambda_1=-\varepsilon$, $\lambda_2=-\xi$.
At this fixed point, all the interactions are irrelevant and the model is free (Gaussian).
It is IR attractive for $\varepsilon<0$, $\eta<0$, and $\xi<0$.

The point (2a) with the coordinates $w'^*=0$, $g^*=32\varepsilon/9$; the corresponding eigenvalues are $\lambda_1=2\varepsilon/3-\xi$ and $\lambda_2=\varepsilon$. 
This point is IR attractive in the area $\eta<0$, $\varepsilon>0$, $\xi<2\varepsilon/3$. Since $w^*=0$, 
turbulent motion of the environment is irrelevant in this regime and IR behavior of the model is completely determined by the universality class of the original Hwa--Kardar model~(\ref{eq1}).

The point (3a) with the coordinates $w'^*=2\xi$, $g^*=0$; the corresponding eigenvalues are $\lambda_1=3\xi/2-\varepsilon$ and $\lambda_2=\xi$. This point is IR attractive when $\eta<0$, $\xi>0$, $\xi>2\varepsilon/3$. Since $g^*=0$, 
the nonlinear term in equation~(\ref{eq1}) is IR irrelevant in the sense of Wilson and does not affect the leading terms of the IR asymptotic behavior.

It is left to note that the functions $\beta_g$ and $\beta_{w'}$ become proportional when $\xi=2\varepsilon/3$, which leads to a straight line of the fixed points in the plane $(g, w')$ (or, in other words, to a single degenerate fixed point), 
with both $g^*\neq0$ and $w'^*\neq0$. They are arbitrary; only a certain combination can be found in a unique way from the system~\eqref{BM1}. As a consequence, one of the eigenvalues is equal to zero.

Now let us turn to the rapid-change case of $\alpha^*\to\infty$, $\lambda_{\alpha}=\eta$. 
As before, it is convenient to pass to new variables which are finite when $\alpha\to\infty$, namely, $x=1/\alpha$ and $w''=w/\alpha^2$. The corresponding $\beta$ functions are 
\begin{eqnarray}
\nonumber
\beta_x&=&x\,\eta; \\ 
\beta_{w''}&=&w''\left(-\xi+\eta+ \frac{w''}{2}+\frac{3}{16}\,g\right).
\label{BM1b}
\end{eqnarray}
Thus, all these fixed points have the coordinate $x^*=0$ and the corresponding eigenvalue is $\lambda_{x}=\eta$. From Eq.~\eqref{BM1b} it follows that the set of the fixed points (1b), (2b), and (3b) at $x^*=0$ is completely the same as the set (1a), (2a), and (3a) for $\alpha^*=0$ after the replacement 
$\xi\to\widetilde\xi=\xi-\eta$ is made in the previously obtained expressions. 
The relation similar to Eq.~\eqref{relation} is also true, thus, there is a line of fixed points for the case $x^*=0$, too. Now the corresponding relation for the exponents reads $\xi-\eta=2\varepsilon/3$. 

It remains to consider the case $\eta=0$. Then the function $\beta_{\alpha}$ vanishes for any given $\alpha$, with the corresponding eigenvalue
$\lambda_{\alpha}=-\eta=0$.
The nontrivial fixed point still exists if $\xi=2\varepsilon/3$. In this case it is determined by the condition $\gamma_{\nu_{\|}}^*=\xi$, see Eqs.~\eqref{betagw2} and~\eqref{BM1}. Herewith, the parameters $g^*$, $w^*$ and $\alpha^*$ cannot be determined independently.

\begin{figure}[t]
\includegraphics[width=8cm]{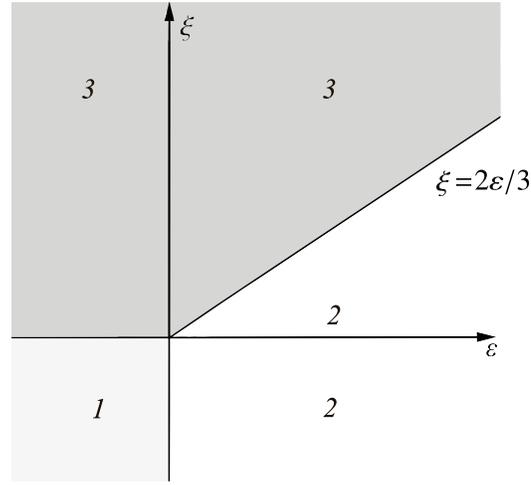}
\caption{Regions of stability of the fixed points for Model~1 at $\alpha^*=0$, $\eta<0$ in the plane $(\varepsilon,\xi)$. Each sector corresponds to the values of the parameters for which one of the points (1a), (2a), or (3a) is IR attractive. }
\label{fig:1}
\end{figure}

\begin{figure}[b]
\includegraphics[width=8cm]{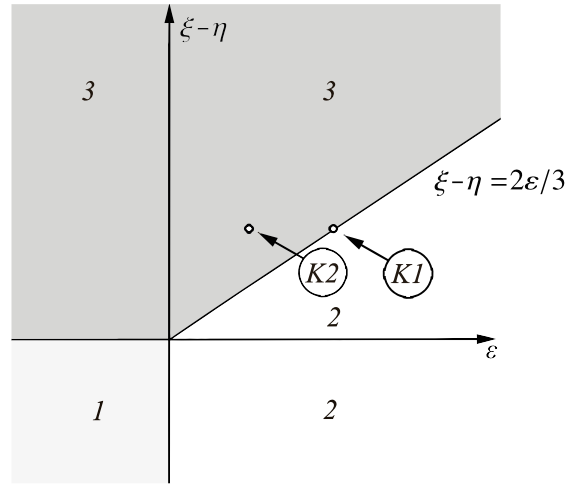}
\caption{Regions of stability of the fixed points for Model~1 at $\alpha^*\to\infty$, $\eta>0$ in the plane $(\varepsilon,\,\xi-\eta)$ which is used instead of the space $(\varepsilon,\xi)$ in Fig.~\ref{fig:1}.
Each sector corresponds to the values of the parameters for which one of the points (1b), (2b), or (3b) is IR attractive. The positions of the fixed points with the Kolmogorov value $\xi-\eta=4/3$ are shown for $d=2$ ($K1$) and $d=3$ ($K2$).}
\label{fig:2}
\end{figure}

The general pattern of stability is shown in Fig.~\ref{fig:1}. The straight lines denote borders of the stability regions (areas where the points are IR attractive); the white color and different types of grey color denote each region.
The Kolmogorov values of the exponents
$\xi=8/3$, $\eta=4/3$ lie either in the stability region of the fixed point (2b) (universality class of the Hwa--Kardar model) or in the stability region of the fixed point (3b) (universality class of the rapid change ensemble) depending on whether $\varepsilon$ is bigger or smaller than $2$, respectively. This 
 means, that if $d\geq3$, the point (3b) corresponds to the Kolmogorov values; if $d=2$, Kolmogorov values lie on the borderline between two regions (see Fig.~\ref{fig:2}).
 
%%%%%%%%%%%%%%%%%%%%%%%%%%%%%%%%%%%%%%%%%%%%%%%%%%%%%%%%%%%%%%%%%%%%%%%%%%%%%%%%%%%%%%%%%%%%%%%%%%%%%%%%%

\subsection{Model 2: the model with the spatially quenched noise}

%%%%%%%%%%%%%%%%%%%%%%%%%%%%%%%%%%%%%%%%%%%%%%%%%%%%%%%%%%%%%%%%%%%%%%%%%%%%%%%%%%%%%%%%%%%%%%%%%%%%%%%%%%

Unlike the  Model~1, the Model~2 involves several renormalization constants, namely three. This leads to drastic difference between the patterns of the RG fixed points, and, hence, possible IR scaling regimes.

The $\beta$ functions now read (we recall that $\tilde\varepsilon=6-d$)
\begin{eqnarray}
\beta_w=-w\left(\xi+\eta+\gamma_w\right), \quad
\beta_g=-g\left(\tilde\varepsilon+\gamma_g\right), \quad
\beta_{\alpha}=-\alpha\eta,
\label{stbetagw}
\end{eqnarray}
with two independent anomalous dimensions $\gamma_g$ and $\gamma_w$.

From Eqs.~\eqref{ZS1} and~\eqref{ZS2} it follows that in the one-loop approximation they have the form
\begin{eqnarray}
\gamma_w=-\frac{w}{2\alpha\,(1+\alpha)}-g, \quad
\gamma_g=-\frac{3w}{4\alpha(1+\alpha)}-\frac{4g}{3}.
\label{stbetagw2}
\end{eqnarray}
Thus, the system of $\beta$ functions analogous to~\eqref{BM1} now reads 
\begin{eqnarray} \nonumber
\beta_w&=&w\left[-\xi-\eta+\frac{w}{2\alpha(1+\alpha)}+
g\right];\\ \nonumber
\beta_g&=&g\left[-\tilde\varepsilon+\frac{3w}{4\alpha(1+\alpha)}+\frac{4g}{3}\right];\\ 
\beta_{\alpha}&=&-\alpha\eta. 
\label{BM2}
\end{eqnarray}

The following stage is the analysis of the fixed points and their stability regions. It reveals essential difference between the patterns of IR asymptotic regimes in the two models.

So far, the RG analysis of the two models was almost identical.
Indeed, they have the same number of coupling constants and they are both multiplicatively renormalizable (although with different number of needed counterterms). 
They have different logarithmic dimensions, but it is possible to perform the RG analysis near the corresponding logarithmic dimension and then return to the physical values by appropriate choice of $\varepsilon$ or $\tilde\varepsilon$.
What is more, the set of the $\beta$ functions~\eqref{BM2} looks very similar to the set~\eqref{BM1}. Nevertheless, the analysis of the expressions~\eqref{BM2} leads to essentially different pattern of the RG flows than those obtained in the previous subsection. 

Let us discuss the fixed points for the Model~2. As before, there are only two possibilities for $\alpha$: $\alpha^*=0$ and $1/\alpha^*\to 0$. Thus, the first case  can be
IR attractive only if $\eta<0$, while the second one can be attractive if $\eta>0$. In this sense, the situation is completely the same as that for the Model~1, including the substitution $\xi\to{\widetilde\xi}=\xi-\eta$ in the obtained expressions. For this reason, below we will consider in detail only the case $\alpha^*=0$. 

After the suitable replacement $w' \to w/\alpha$, the new 
$\beta$ function reads
\begin{equation}
\beta_{w'}=w'\left(-\xi+\frac{w'}{2}+g\right).
\label{BM2a}
\end{equation}
The system~\eqref{BM2}~-- \eqref{BM2a} possess four different solutions. The fixed points (1a), (2a), and (3a) are the natural counterparts to the same-denoted points 
in Model~1; the point (4a) has no analog and is completely new. 

The point (1a) has the coordinates $w'^*=0$, $g^*=0$. The eigenvalues of the matrix~\eqref{Omega} are $\lambda_1=-\tilde\varepsilon$, $\lambda_2=-\xi$. This Gaussian point is IR attractive for $\tilde\varepsilon<0$, $\eta<0$, $\xi<0$.

The point (2a) has the the coordinates $w'^*=0$, $g^*=3\tilde\varepsilon/4$. The corresponding eigenvalues are $\lambda_1=3\tilde\varepsilon/4-\xi$ and $\lambda_2=\tilde\varepsilon$.
The point is related to the universality class of the pure Hwa--Kardar model without turbulent advection and is IR attractive for $\eta<0$, $\tilde\varepsilon>0$, $\xi<3\tilde\varepsilon/4$.

The point (3a) has the coordinates  $w'^*=2\xi$ and $g^*=0$. The corresponding eigenvalues are $\lambda_1=3\xi/2-\tilde\varepsilon$ and $\lambda_2=\xi$.
This point corresponds to the regime in which the 
nonlinearity of the Hwa--Kardar equation is irrelevant; it is IR attractive if $\eta<0$, $\xi>0$, $\tilde\varepsilon<3\xi/2$.

The completely new point (4a) has the coordinates
\begin{equation}
w'^*=12\tilde\varepsilon-16\xi,\quad g^*=9\xi-6\tilde\varepsilon.
\end{equation}
Its eigenvalues read
\begin{equation}
\lambda_{1,2}=-\tilde\varepsilon+2\xi\pm\sqrt{-5\tilde\varepsilon^2+13\xi\tilde\varepsilon-8\xi^2}.
\label{Root}
\end{equation}
The analysis of Eqs.~\eqref{Root} reveals two possible cases: the square root is either fully real (the case A) or complex with both real and imaginary parts (the case B). 
The presence of imaginary parts in eigenvalues means that if this point is IR attractive (i.e., the real parts of the eigenvalues are positive), it is a spiral attractor instead of a simple node attractor.

The case A corresponds to the two areas of the values of the system parameters: $\tilde\varepsilon>0$, $5\tilde\varepsilon/8<\xi<2\tilde\varepsilon/3$, and $\tilde\varepsilon>0$, $3\tilde\varepsilon/4<\xi<\tilde\varepsilon$. 
The case B also corresponds to the two areas: $\tilde\varepsilon>0$, $\tilde\varepsilon/2<\xi<5\tilde\varepsilon/8$ and the large area which is parameterized by conditions $\tilde\varepsilon>0$, $\xi>\tilde\varepsilon$ and $\tilde\varepsilon<0$, $\xi>\tilde\varepsilon/2$.
It is very interesting that we see a gap in the region of stability of these points; moreover, it is intriguing that the area related to the node attractor  lies beyond the area of spiral attractor.

Another surprising fact is that the borders of the gap in the stability region of the point~(4a) completely coincide with the upper and lower borders of stability regions of the points (2a) and (3a) which are defined by their own (independent) eigenvalues. It is also very interesting that even the gap of the point (4a) is a stability region of two fixed points. 

Moreover, the stability region of one of the nontrivial fixed points lies in the area $\tilde\varepsilon<0$, $\xi<0$. We have never before met a system with such a feature. 

\begin{figure}[t]
\includegraphics[width=8cm]{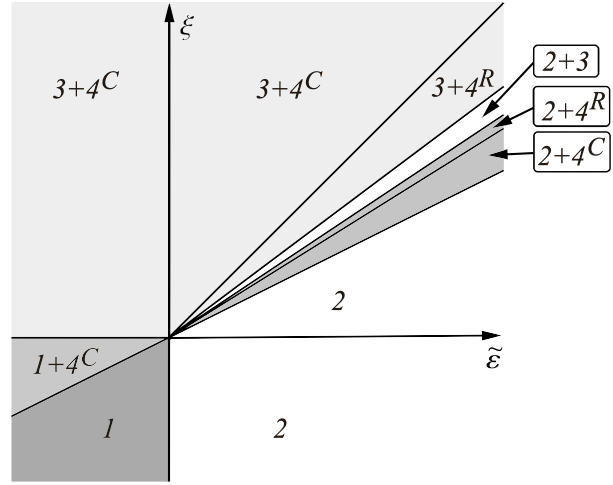}
\caption{Regions of stability of the fixed points for Model~2 at $\alpha^*=0$, $\eta<0$ in the plane $(\tilde\varepsilon,\xi)$. Each sector corresponds to the values of the parameters for which some of the points (1a), (2a), (3a), or (4a) are IR attractive. Notations like $(2+4)$ stand for the sectors where some regions overlap. Designation $4^R$ corresponds to the part of the 
stability region of the point (4a)  where it is a simple node. Designation $4^C$ corresponds to the part where it is a spiral fixed point (an attractive focus).}
\label{fig:stat}
\end{figure}

The general pattern of stability is shown in Fig.~\ref{fig:stat}. The straight lines denote borders of the stability regions (areas where the points are IR attractive); the white color and different types of grey color denote each region.
The subscripts ``R'' and ``C'' near the point (4) denote the type of the root in Eqs.~\eqref{Root} and, therefore, the type of the attractor. Designation $4^R$ corresponds to the node attractor, designation $4^C$ corresponds to the spiral attractor. The points (1), (2), and (3) do not have such variants and, therefore, have no subscripts. 

The main reason for this drastic difference between the results obtained for Model~1 and Model~2 is the absence of the relation like~\eqref{relation} in Model~2. This is, in its turn, a consequence of there being more than one independent renormalization constant $Z$ in the model. It opens the possibility for a fixed point with both $g^*\neq0$ and $w'^*\neq0$ to exist. Therefore, it was natural to expect the set of fixed points in Model~2 to be more interesting and rich than the set in Model~1. 
Nevertheless, one could hardly expect the obtained picture of fixed points to be so complicated and to consist of points with overlapping stability regions that 
even have gaps in them.\footnote{It is important to note that not only stability region of the ``new'' point 4 intersects with the regions of other fixed points but that there are also overlaps between stability regions of the points 2 and 3. In Model~1 stability regions of the similar points have no such overlaps.} 
This result is really surprising; nothing at the start of the analysis of the system~\eqref{BM2} indicated that the result was to be expected. 

Overlaps between the stability regions of different fixed points have  important implication for the universality of system's asymptotic behavior. Universality means that the behavior depends only on the global characteristics of the system like spatial dimension $d$ and values of $\xi$ and $\eta$. But if several fixed points share a stability region, then the RG flow may reach either one of them depending on the initial values of the coupling constants. This dependence can be interpreted as a loss of universality (universality violation).

\begin{figure}[t]
\includegraphics[width=8cm]{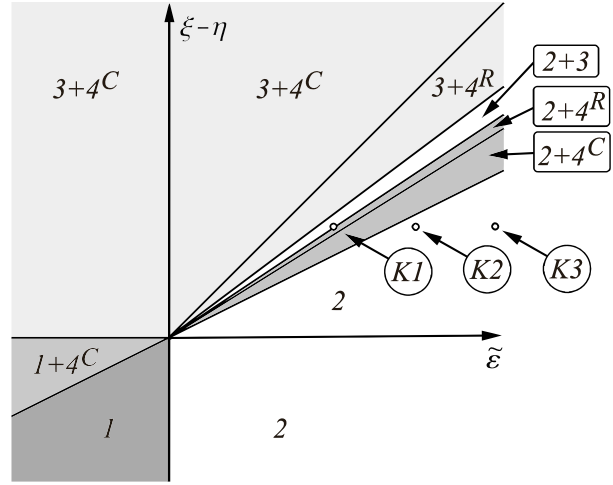}
\caption{Regions of stability of the fixed points for Model~2 at $\alpha^*\to\infty$, $\eta>0$ in the plane $(\tilde\varepsilon,\xi-\eta)$ which is used instead of the space $(\tilde\varepsilon,\xi)$ in Fig.~\ref{fig:stat}.
Each sector correspond to the values of the parameters for which some of the points (1b), (2b), (3b), or (4b) are IR attractive. 
Similarly to Fig.~\ref{fig:stat}, notations like $(2+3)$ stand for the sectors where the stability regions overlap. The positions of the fixed points with the Kolmogorov value $\xi-\eta=4/3$ are shown for $d=2$ ($K1$), $d=3$ ($K2$), and $d=4$ ($K3$).}
\label{fig:stat3}
\end{figure}

The Kolmogorov values of the exponents are 
$\xi=8/3$, $\eta=4/3$, and the corresponding stability region depends on $\tilde\varepsilon$. As such, the Kolmogorov values may relate to either of the following regions: the area where both fixed points (3b) and (4b) are IR attractive, the area where both (2b) and (3b) are IR attractive, the area where both (2b) and (4b) are IR attractive, or the area where only (2b) is IR attractive. However, at $\eta=4/3$ the wedges in Fig.~\ref{fig:stat} are very small and most of the regions are unattainable for integer values of $d$. The result of this is that for every $\tilde\varepsilon>8/3$, i.e., for $d\leq3$, 
the Kolmogorov values belong to the stability region of the fixed point (2b). If $\tilde\varepsilon=2$, i.e., if $d=4$, the Kolmogorov values lie on the borderline between two  regions: the region where both fixed points (2b) and (4b) are IR attractive and the region where both points (2b) and (3b) are IR attractive (see Fig.~\ref{fig:stat3}).
\pagebreak

\begin{figure}[b]
\includegraphics[width=17cm]{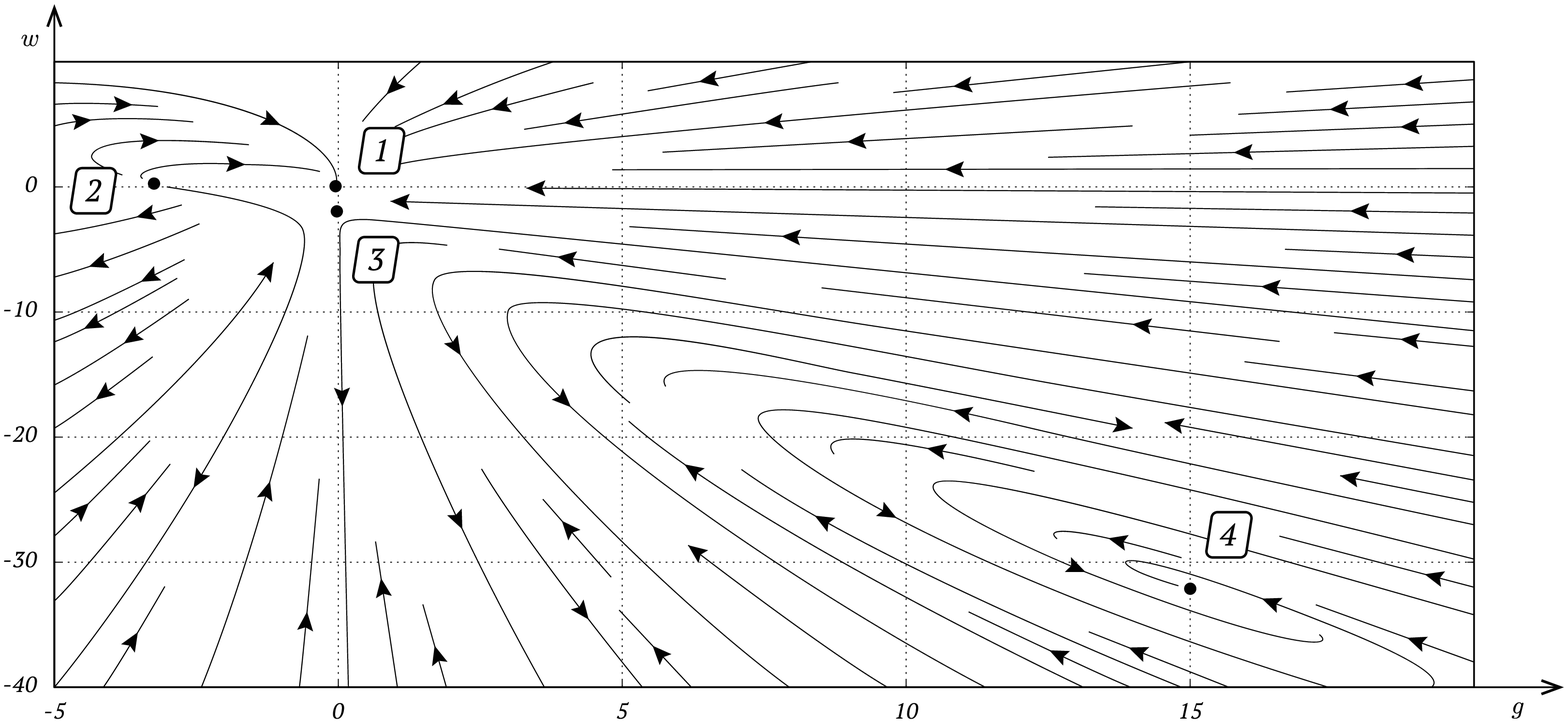}
\caption{RG flow in Model 2 for $\alpha=0$, ${\tilde \varepsilon}=-4$, $\xi=-1$. The numbers indicate the fixed points (1a), (2a), (3a), and (4a). The fixed point (1a) is a node attractor while the fixed point (4a) is a spiral attractor.}
\label{fig4}
\end{figure}

\begin{figure}[t]
\includegraphics[width=16cm]{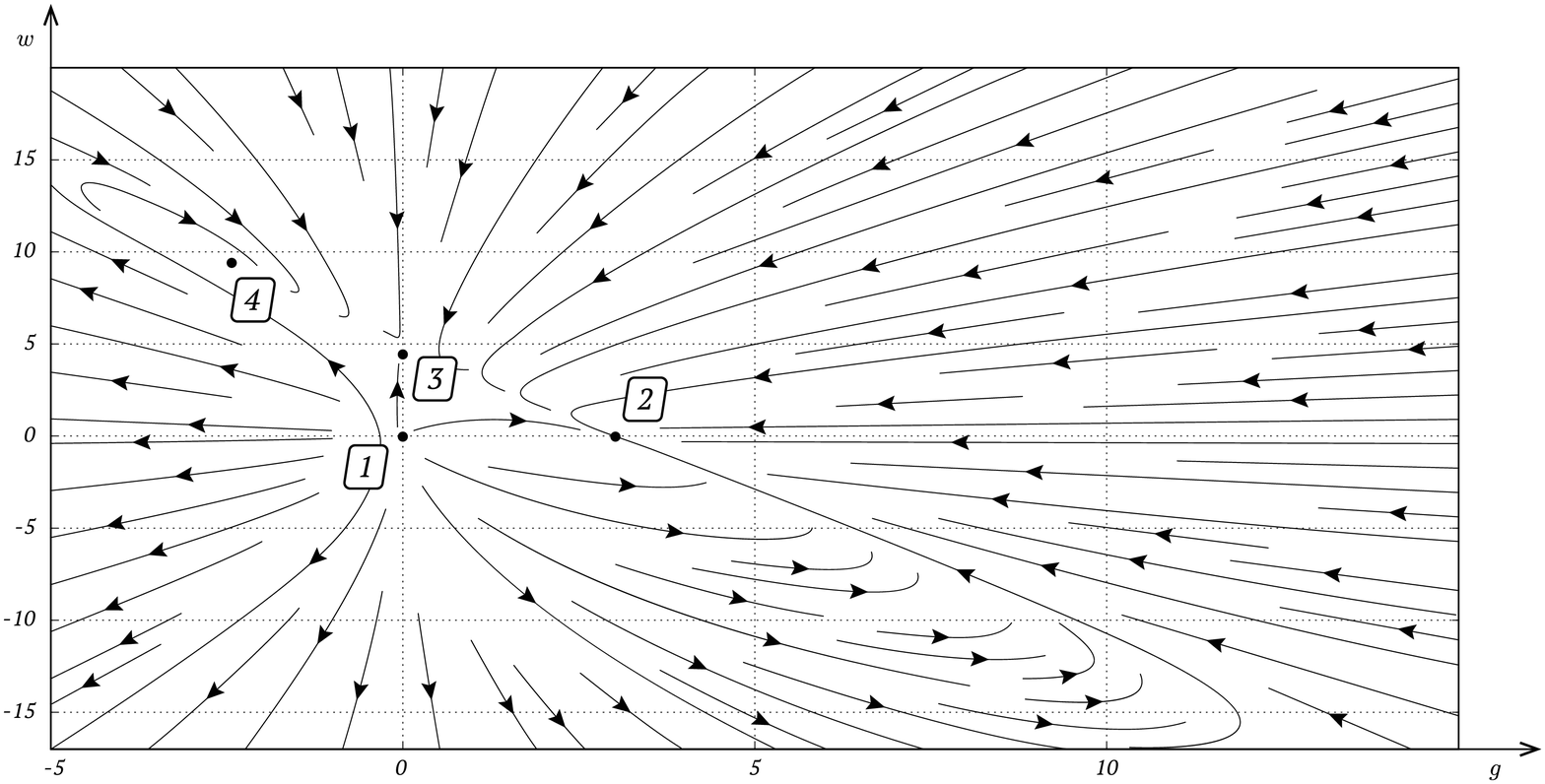}
\caption{RG flow in Model 2 for $\alpha=0$, ${\tilde \varepsilon}=4$, $\xi=2.4$. The numbers indicate the fixed points (1a), (2a), (3a), and (4a). The fixed point (2a) is a node attractor while the fixed point (4a) is a spiral attractor.}
\label{fig5}
\end{figure}

The obtained results look rather complicated, so we used dynamical equations~\eqref{eq:invariant_chrg} for direct numerical simulation of the RG flows. We carefully checked  each stability region and obtained a full agreement with the analytical analysis.
In Figs.~\ref{fig4} and~\ref{fig5} two sample RG flows are presented for interesting values  $\alpha=0$, ${\tilde \varepsilon}=-4$, $\xi=-1$ and $\alpha=0$, ${\tilde \varepsilon}=4$, $\xi=2.4$,
where two IR attractive fixed points (a node and a focus) exist simultaneously.
The arrows on the lines designate a direction towards the IR limit $s=k/\mu\to0$.
\pagebreak 

%%%%%%%%%%%%%%%%%%%%%%%%%%%%%%%%%%%%%%%%%%%%%%%%%%%%%%%%%%%%%%%%%%%%%%%%%%%%%%%%%%%%%%%%%%%%%%%%%%%%%%%%%%%%%%%%%%%%%%%%%%%%%%%%%%%%%%%%%%%%%%%%%%%%%%%%%%%%%%%%%%%%%%%%%%%%%%%%%%%%%%%%%%

\section{Critical scaling and critical dimensions.} 
\label{sec:DimeNS1}

%%%%%%%%%%%%%%%%%%%%%%%%%%%%%%%%%%%%%%%%%%%%%%%%%%%%%%%%%%%%%%%%%%%%%%%%%%%%%%%%%%%%%%%%%%%%%%%%%%%%%%%%%%%%%%%%%%%%%%%%%%%%%%%%%%%%%%%%%%%%%%%%%%%%%%%%%%%%%%%%%%%%%%%%%%%%%%%%%%%%%%%%%%

The renormalized Green functions $G^{R}$ satisfy RG equation in the leading order of IR asymptotic behavior when the substitution $q\to q^*$ is made. This is a consequence of expressions~\eqref{Asym}; $q$ is the set of three coupling constants. As a result, the RG equation reads
\begin{equation}
\left(
{\cal D}_{\mu} - \sum_{i}\gamma_{i}^{*}{\cal D}_{i}
+ \sum_{\Phi} N_{\Phi}\gamma_{\Phi}^{*}  \right) \, G^{R} = 0,
\label{RGF}
\end{equation}
where $\gamma_{\Phi}$ and $\gamma_{i}$ are anomalous dimensions of the fields and the parameters that require renormalization, respectively. Since the values of anomalous dimensions at a fixed point $\gamma^*_{\Phi}$ and $\gamma^*_{i}$ are constants, equation~\eqref{RGF} is a differential equation with constant coefficients and, therefore, is an equation of the same type as differential equations for canonical scale invariance. 
Solution of the system of equations that includes Eq.~\eqref{RGF} together with the equations for  canonical scale invariance gives us 
critical dimension $\Delta_{F}$ of an IR relevant
quantity $F$ (a field or a parameter), see~\cite{Book3}. 
Since we have two spatial (momentum) scales while for both models renormalization constants $Z_{\nu_{\parallel}}\neq0$ and $Z_{\nu_{\bot}}=1$, this dimension reads~\cite{Alexa,shark}
\begin{equation}
\Delta_{F} = d_{F}^{\bot} + \Delta_{\parallel} d_{F}^{\parallel}  +
\Delta_{\omega}d_{F}^{\omega} + \gamma_{F}^{*}; \quad \Delta_{\parallel} = 1 + \gamma_{\nu_{\parallel}}^{*}/2, \quad \Delta_{\omega}=2.
\label{dim}
\end{equation}
The factor $1/2$ in expression for $\Delta_{\parallel}$ is due to canonical dimension $d^\parallel_{\nu_\parallel}=-2$; since $\gamma_{\nu_{\bot}}=0$, the critical dimension of the frequency~$\Delta_\omega$ is simply equal to 2. 
As usual, $d_{F}$ are canonical dimensions (see Tables~\ref{t1} and~\ref{t2}) and 
$\gamma_{F}^{*}$ is corresponding anomalous
dimension taken at the fixed point. The normalization condition $\Delta_{\bot}=1$ is used. 

Depending on the  values of $\eta$, $\xi$, and $\varepsilon$ or $\tilde\varepsilon$, RG flow reaches certain fixed point. Substitution of the fixed point coordinates ($g^*$, $w^*$, and $\alpha^*$) into Eqs.~\eqref{dim} leads to expressions for the  critical dimensions that correspond to the possible scaling regimes of the system. 

Calculation of critical dimensions is the final goal of the general scheme: 
they appear in the pair correlation function of the field $h$ in the following way
\begin{equation}
\langle h(t,{\bm x})\, h(0,{\bm 0}) \rangle \simeq
r_{\bot}^{-2\Delta_{h}}\, {\cal F} \left(t/ r_{\bot}^{\Delta_{\omega}},
r_{\parallel}/ r_{\bot}^{\Delta_{\parallel}} \right)
\label{dimS}
\end{equation}
and allow direct comparison with experiments.
Here $r_{\bot}=|{\bm x}_{\bot}|$, $r_{\parallel}=x_{\parallel}$, and
${\cal F}$ is a scaling function of critically dimensionless arguments.

%%%%%%%%%%%%%%%%%%%%%%%%%%%%%%%%%%%%%%%%%%%%%%%%%%%%%%%%%%%%%%%%%%%%%%%%%%%%%%%%%%%%%%%%%%%%%%%%%%%%%%%%%

\subsection{Model 1: the model with the white noise}

%%%%%%%%%%%%%%%%%%%%%%%%%%%%%%%%%%%%%%%%%%%%%%%%%%%%%%%%%%%%%%%%%%%%%%%%%%%%%%%%%%%%%%%%%%%%%%%%%%%%%%%%%

The functions  $\beta_g$ and $\beta_w$ involve the same anomalous dimension $\gamma_{\nu_{\|}}$ [see relation~\eqref{relation}] in Model~1, so the value of $\gamma^*_{\nu_{\|}}$ and the critical dimensions are found exactly despite the fact that coordinates of the fixed points are found only in one-loop approximation. 
This nontrivial fact reminds of a similar observation in the stochastic NS equation~\cite{DM,UFN,Red},
where all the anomalous dimensions can be found exactly without any practical calculation of the renormalization constants. 

Indeed, if one of the coupling constants (say, $w$) is necessary equal to zero at fixed point, it follows from Eqs.~\eqref{betagw} that $\gamma^*_{\nu_{\|}}=2\varepsilon/3$.
Since Eqs.~\eqref{betagw} follow directly from the definitions of $\beta$ and $\gamma$ functions, they are exact. Therefore, the value of $\gamma^*_{\nu_{\|}}$ obtained above is also exact. 
Nevertheless, it is necessary to calculate Feynman graphs to check the stability regions of different fixed points (i.e., to find the derivatives of $\beta$ functions at the fixed points).

Critical dimensions for the trivial points (1a) and (1b) coincide with each other and read
\begin{equation}
\Delta_{h'}=d-1=3-\varepsilon, \quad \Delta_h=\Delta_v=\Delta_{\parallel}=1.
\end{equation}
Critical dimensions for the fixed points (2a) and (2b) also coincide with each other and read 
\begin{equation}
\Delta_{h'}=3-\frac{\varepsilon}{3}, \quad \Delta_h=\Delta_v=1-\frac{\varepsilon}{3},\quad \Delta_{\parallel}=1+\frac{\varepsilon}{3}.
\end{equation}
Critical dimensions for the fixed points (3a) and (3b) are not the same and read
\begin{eqnarray} 
\label{dim3a}
\Delta_{h'}&=&3-\varepsilon+\xi, \quad \Delta_h=\Delta_v=1-\frac{\xi}{2}, \quad \Delta_{\parallel}=1+\frac{\xi}{2} \quad \text{for the point (3a)};\\
\label{dim3b}
\Delta_{h'}&=&3-\varepsilon+\xi-\eta, \quad \Delta_h=\Delta_v=1-\frac{\xi-\eta}{2}, \quad \Delta_{\parallel}=1+\frac{\xi-\eta}{2} \quad \text{for the point (3b)}.
\end{eqnarray}
As it should be, the results for the fixed point (2b) agree with those obtained in~\cite{HK}.\footnote{One has to identify
$z=\Delta_{\omega}/\Delta_{\parallel}$, $\zeta=1/\Delta_{\parallel}$, and $\chi = -\Delta_{h}/\Delta_{\parallel}$.}
The results for all the three points (1b), (2b), and (3b) that correspond to the rapid-change velocity ensemble agree with those obtained in~\cite{AK1}.\footnote{Here, one has to identify
$\xi$ from~\cite{AK1} with $\xi -\eta$ in Eqs.~\eqref{dim3b}; moreover, there are misprints in~\cite{AK1} in expressions for $\Delta_{h'}$.}

One important remark is in order here. For the Kazantsev--Kraichnan rapid-change ensemble the diagram $D_1$ (see Appendix~\ref{App1} for details) involves an indeterminacy, which in~\cite{AK1} was tacitly understood as
$1/(2\pi)\int d\omega/[-{i}\omega+\epsilon(k)]=1/2$; 
cf. also \cite{shark,Ito5}. This resolution is 
justified by the physical meaning of the pair correlation function (\ref{veloc1})~-- (\ref{veloc2}) and in the theory of stochastic equations is known as the Stratonovich prescription \cite{Ito1,Ito2}. As applied to field-theoretic formulations,
the most detailed and comprehensive discussion of the issue is given in~\cite{Ito23,Ito3,Ito4}. Thus, the model studied in~\cite{AK1} is indeed a special case of the present model in the limit $\alpha_0 \to \infty$ at the fixed $B_0 / \alpha_0^2$; cf.~\cite{shark}. 

%%%%%%%%%%%%%%%%%%%%%%%%%%%%%%%%%%%%%%%%%%%%%%%%%%%%%%%%%%%%%%%%%%%%%%%%%%%%%%%%%%%%%%%%%%%%%%%%%%%%%%%%%

\subsection{Model 2: the model with spatially quenched noise}

%%%%%%%%%%%%%%%%%%%%%%%%%%%%%%%%%%%%%%%%%%%%%%%%%%%%%%%%%%%%%%%%%%%%%%%%%%%%%%%%%%%%%%%%%%%%%%%%%%%%%%%%%

In Model~2 there is no exact relation between $\beta_g$ and $\beta_w$; therefore, all of the results (both coordinates of the fixed points and the critical dimensions) imply corrections in $\tilde\varepsilon$, $\xi$, and $\eta$ of second order and higher (except the critical dimensions for the points (3a) and (3b), see below). 
Moreover, since the fields $h$, $h'$, and $v$ have nontrivial renormalization constants, now we should take into account corresponding anomalous dimensions. 
From Eqs.~\eqref{ZS1} and~\eqref{ZS2} it follows that they read
\begin{equation}
\gamma_h=\gamma_v=-\frac{g}{6}, \quad \gamma_{h'}=\frac{g}{6}.
\label{gammas}
\end{equation}

Critical dimensions for the trivial points (1a) and (1b) are still very simple and read
\begin{equation}
\Delta_{h'}=5-\tilde\varepsilon, \quad \Delta_h=\Delta_v=\Delta_{\parallel}=1.
\end{equation}
Critical dimensions for the points (2a) and (2b) are 
\begin{equation}
\Delta_{h'}=5-\frac{3\tilde\varepsilon}{8}, \quad \Delta_h=\Delta_v=1-\frac{3\tilde\varepsilon}{8}, \quad \Delta_{\parallel}=1+\frac{\tilde\varepsilon}{4}.
\end{equation}

Since the differences between Eqs.~\eqref{BM2} (for Model 2) and Eqs.~\eqref{BM1} (for Model 1) are presented only in the parts that contain the coupling constant $g$, the coordinates of the fixed points (3a) and (3b) (which satisfy the case $g^*=0$) coincide for both models.
Owning to this fact and Eqs.~\eqref{gammas}, the only difference in the critical dimensions for the points (3) in both models is due to their canonical parts which are connected with the logarithmic dimensions $d=6$ for Model~2 and $d=4$ for Model~1. 
Thus, the critical dimensions for the points (3a) and (3b) are
\begin{eqnarray} 
\label{dim3a2}
\Delta_{h'}&=&5-\tilde\varepsilon+\xi, \quad \Delta_h=\Delta_v=1-\frac{\xi}{2}, \quad \Delta_{\parallel}=1+\frac{\xi}{2} \quad \text{for the point (3a)};\\
\label{dim3b2}
\Delta_{h'}&=&5-\tilde\varepsilon+\xi-\eta, \quad \Delta_h=\Delta_v=1-\frac{\xi-\eta}{2}, \quad \Delta_{\parallel}=1+\frac{\xi-\eta}{2} \quad \text{for the point (3b)}.
\end{eqnarray}
Like in Model~1, these results are exact; see Appendix B for details.

The critical dimensions for the fully nontrivial points (4a) and (4b) read:
\begin{eqnarray} 
\label{dim3a3}
\Delta_{h'}&=&5-\frac{\xi}{2}, \quad \Delta_h=\Delta_v=1-\frac{\xi}{2}, \quad \Delta_{\parallel}=1+\tilde\varepsilon-\xi \quad \text{for the point (4a)};\\
\label{dim3b3}
\Delta_{h'}&=&5-\frac{\xi-\eta}{2}, \quad \Delta_h=\Delta_v=1-\frac{\xi-\eta}{2}, \quad \Delta_{\parallel}=1+\tilde\varepsilon-\xi+\eta \quad \text{for the point (4b)}.
\end{eqnarray}

The above expressions show that the critical dimensions of the fields $h$ and $v$ coincide with each other, and, what is more, coincide for the points (3) and (4) [a and b, respectively]. This coincidence looks intriguing for two reasons: first, the answers for fixed points~(3) are exact while the answers for points (4) admit higher-order corrections in $\xi^2$ and $\eta^2$.
Second, in contrast to the coincidence of critical dimensions for points (3) in the two different models [see Eqs.~\eqref{dim3a}~-- \eqref{dim3b} and \eqref{dim3a2}~-- \eqref{dim3b2}], the algebraic manipulations that lead to the same results for points (3) and (4) in Model~2 are essentially different. Therefore, the equality between the critical dimensions at different points may be both just an artifact of the one-loop approximation, or may be manifestation of some underlying physics.

%%%%%%%%%%%%%%%%%%%%%%%%%%%%%%%%%%%%%%%%%%%%%%%%%%%%%%%%%%%%%%%%%%%%%%%%%%%%%%%%%%%%%%%%%%%%%%%%%%%%%%%%%%%%%%%%%%%%%%%%%%%%%%%%%%%%%%%%%%%%%%%%%%%%%%%%%%%%%%%%%%%%%%%%%%%%%%%%%%%%%%%%%%

\section{Conclusion} 
\label{sec:Conc}

%%%%%%%%%%%%%%%%%%%%%%%%%%%%%%%%%%%%%%%%%%%%%%%%%%%%%%%%%%%%%%%%%%%%%%%%%%%%%%%%%%%%%%%%%%%%%%%%%%%%%%%%%%%%%%%%%%%%%%%%%%%%%%%%%%%%%%%%%%%%%%%%%%%%%%%%%%%%%%%%%%%%%%%%%%%%%%%%%%%%%%%%%%

In this paper we apply the field theoretic renormalization group to two models of self-organized nearly-critical systems of statistical physics. In the spirit of Hwa and Kardar, both problems are described by the continuous (coarse-grained)
stochastic differential equations, subjected to a random noise.
In the first case, the random noise is taken to be white in time and in space, while in the second case the noise is ``spatially quenched,'' that is, white in space and time-independent.

Both models are intended to describe the effects of turbulent environment on the critical behavior of the initial systems. The environment motion is described by the $d$-dimensional generalization of the Avellaneda--Majda ensemble~\cite{AM,AM1} with a finite correlation time and strong anisotropy,
conformed with that of the Hwa--Kardar model.

The quantities of interest are the critical dimensions of the fields and parameters, related to the asymptotic forms of the (measurable) correlation functions like~(\ref{dimS}). Those dimensions are determined by IR attractive fixed points. Thus, our ultimate goal is to identify the sets of fixed points and their regions of IR stability for the models under study. Our analysis shows that, despite the fact that the models look very similar to each other (even in the expressions for the $\beta$ functions), the resulting patterns of the fixed points are drastically different. 

While for the white-noise case the picture obtained is more or less typical for that kind of models, 
i.e., there are several fixed points with neither gaps nor overlaps between their stability regions, the picture for the model with the spatially quenched noise seems to be much more complicated. It contains overlaps between regions of stability of different fixed points. From the physics point of view, this feature may be interpreted as a loss of universality: now the critical dimensions depend not only on global characteristic of the system like space dimension $d$ and the values of the exponents $\xi$ and $\eta$ that characterise the velocity statistics, but also on the initial values of the coupling constants. These initial conditions determine which of the possible fixed points is reached by the RG flow. 

It is interesting that spatially quenched noise is widely used in various models but does not lead to such complicated behavior as a rule. Moreover, the stochastic Hwa--Kardar equation with the spatially quenched noise without the turbulent field ${\bm v}$ does not display such interesting properties; see~\cite{Stat}. Instead, this choice of random noise only shifts the logarithmic dimension of the model, while in the present case we observe different behavior with gaps and overlaps between the stability regions.

The stability regions of the fully nontrivial points (4a) and (4b) for which both the Hwa--Kardar nonlinearity and turbulent advection are relevant appear to have gaps. In some cases~\cite{Lesha3,kap,Menk}, such gaps are completely ``vacant,'' i.e., there are no other IR attractive fixed points for the corresponding values of the parameters. The critical behavior in those cases remains unclear. In the present case, however, the gaps are covered with the stability regions of the points (2a)~-- (3a) and (2b)~-- (3b) respectively, so there is no area with unknown critical behavior.

Although we used the velocity ensemble with finite correlation time,
the possible nontrivial types of the IR behavior reduce to only the two limiting cases: the rapid-change behavior and the frozen (time-independent) case. 
This feature is rather typical, being observed in many different models before; see, e.g.,  \cite{AG2,AG2a,AGM,AGMK,shark,Ant3} and the review paper~\cite{Tomas}. However, fixed points with finite correlation time were encountered in another models due to the presence of compressibility, e.g., in~\cite{Ant2,Ant4}.

Another remarkable fact is that the model with the white noise possess only three IR attractive fixed points with always at least one coupling constant equal to zero, while the model with the spatially quenched noise involves four IR attractive fixed points including the one for which both the couplings are nontrivial.\footnote{To be precise, such point exists for the white-noise case, but only for the special choice $\xi=2\varepsilon/3$.}

We may conclude that the interplay between the random noise with different statistics and the turbulent advection can lead to essentially different patterns of the fixed points, their regions of stability and character of the RG flows.
Comparison of the models with different velocity ensembles may reveal  what ingredients of the formulation of those problems are responsible for such complicated behavior. 

It is also worth noting that the systems of differential RG equations 
like~(\ref{eq:invariant_chrg})
provide real examples of dynamical systems~\cite{Richt,Paladin}. So far, all the known systems like these demonstrated typical kinds of asymptotic behavior: fixed points, manifolds of fixed points, attractive circles. 
Strange attractors were sought, but had never been found~\cite{Niemi}. 
The nonequilibrium models provide a huge variety of dynamical systems and, hopefully, the intrinsic relation between field theoretical models and corresponding dynamical systems (described by the RG equations for invariant variables) will eventually be established.

In particular, it would be especially interesting to consider the Hwa--Kardar model coupled to the velocity field ${\bm v}(x)$ described by the nonlinear stochastic Navier--Stokes equation with various types of external force. In any of those cases, the model acquires additional viscosity coefficient and one more field, namely, the response field ${\bm v}'$, thus, a much more intricate and sophisticated types of asymptotic behavior may be expected. 

Alternatively, it can be instructive to compare models with the same velocity ensemble and different noise statistics.
It is also tempting to consider generalized noise statistics that interpolates between the two limiting cases~(\ref{forceD}) and~(\ref{forceStat}).

However, the RG analysis of such full-scale problems is clearly a very difficult and cumbersome task.
As a preliminary step, the isotropic Kazantsev--Kraichnan ensemble can be employed. The Hwa--Kardar model~(\ref{eq1}) with the white noise~(\ref{forceD}) coupled to that ensemble was studied in~\cite{German}. It was shown that coupling of anisotropic system and isotropic flow leads to rather surprising results: in particular, some dimensionless ratio of diffusivity coefficients acquires nontrivial dimension in certain fixed-point limits.

The work on those more realistic and more complicated systems remains for the future and is partly in progress.

%%%%%%%%%%%%%%%%%%%%%%%%%%%%%%%%%%%%%%%%%%%%%%%%%%%%%%%%%%%%%%%%%%%%%%%%%%%%%%%%%%%%%%%%%%%%%%%%%%%%%%%%%%%%%%%%%%%%%%%%%%%%%%%%%%%%%%%%%%%%%%%%%%%%%%%%%%%%%%%%%%%%%%%%%%%%%%%%%%%%%%%%%%

\section*{Acknowledgments}
The authors are indebted to Elena A. Bragina who courteously helped with preparation of the figures.
The reported study was funded by the Russian Foundation for Basic Research, project number~20-32-70139.
The work by The work by N.V.A. and P.I.K. was also supported by Theoretical Physics and Mathematics Advancement Foundation ``BASIS.''

%%%%%%%%%%%%%%%%%%%%%%%%%%%%%%%%%%%%%%%%%%%%%%%%%%%%%%%%%%%%%%%%%%%%%%%%%%%%%%%%%%%%%%%%%%%%%%%%%%%%%%%%%%%%%%%%%%%%%%%%%%%%%%%%%%%%%%%%%%%%%%%%%%%%%%%%%%%%%%%%%%%%%%%%%%%%%%%%%%%%%%%%%%

%%%%%%%%%%%%%%%%%%%%%%%%%%%%%%%%%%%%%%%%%%%%%%%%%%%%%%%%%%%%%%%%%%%%%%%%%%%%%%%%%%%%%%%%%%%%%%%%%%%%%%%%%%%%%%%%%%%%%%%%%%%%%%%%%%%%%%%%%%%%%%%%%%%%%%%%%%%%%%%%%%%%%%%%%%%%%%%%%%%%%%%%%%%%%%%%%%%%%%%%%%%%%%%%%%%%%%%%%%%%%%%%%%%%%%%%%%%%%%%%%%%%%%%%%%%%%%%%%%%%%%%%%%%%%%%%%%%%%%%%%%%%%%%%%%%%%%%%%%%%%%%%%%%%%%%%%%%%%%%%%%%%%%%%%%%%%%%%%%%%%%%%%%%%%%%%%%%%%%%%%%%%%%%%%%%%%%%%%%%%%%%%%%%%%%%%%%%%%%%%%%%%%%%%%%%%%%%%%%%%%%%%%%%%%%%%%%%%%%%%%%%%%%%%%%%%%%%%%%%%%%%%%%%%%%%%%%%%%%%%%%%%%%%%%%%%%%%%%%%%%%%%

\appendix

%%%%%%%%%%%%%%%%%%%%%%%%%%%%%%%%%%%%%%%%%%%%%%%%%%%%%%%%%%%%%%%%%%%%%%%%%%%%%%%%%%%%%%%%%%%%%%%%%%%%%%%%%%%%%%%%%%%%%%%%%%%%%%%%%%%%%%%%%%%%%%%%%%%%%%%%%%%%%%%%%%%%%%%%%%%%%%%%%%%%%%%%%%

\section{Calculation details of Model~1 }
\label{App1}

%%%%%%%%%%%%%%%%%%%%%%%%%%%%%%%%%%%%%%%%%%%%%%%%%%%%%%%%%%%%%%%%%%%%%%%%%%%%%%%%%%%%%%%%%%%%%%%%%%%%%%%%%%%%%%%%%%%%%%%%%%%%%%%%%%%%%%%%%%%%%%%%%%%%%%%%%%%%%%%%%%%%%%%%

This section contains detailed calculations of the diagrams defining the renormalization
constant $Z_{\nu_\parallel}$ (see Sec.~\ref{App1Z}).
All the calculations are performed in the analytical ($\xi$ and $\eta$) and dimensional ($\varepsilon$) regularization and MS scheme.

Since the model we deal with involves only one divergent function $\langle h'h \rangle_{1-ir}$, there are only two one-loop graphs needed to be calculated: 
\vspace{-.7em}
\begin{eqnarray}
D_1=\raisebox{-3ex}{ \includegraphics[width=3.1truecm]{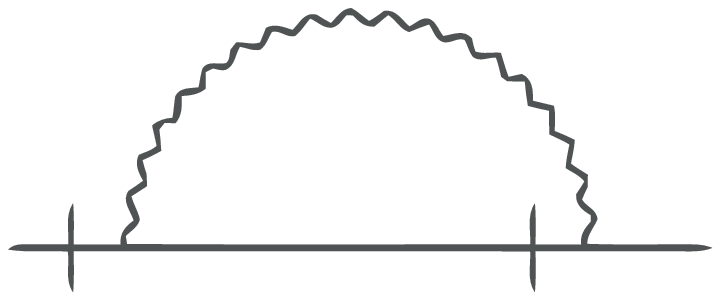}} \quad \text{and} \quad D_2=\raisebox{-3ex}{ \includegraphics[width=3.0truecm]{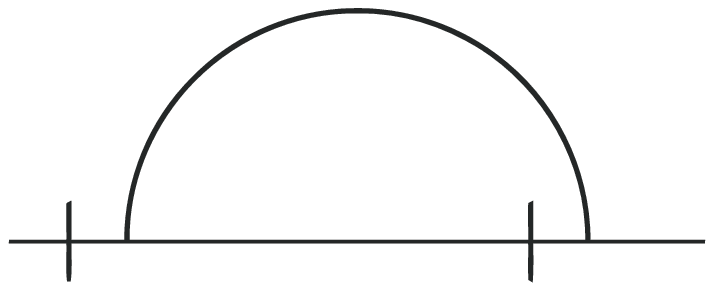}}.
\label{d1d2white}
\end{eqnarray}
Here and below the straight line corresponds to the field $h$, the dashed line corresponds to the field $h'$, and the wave line corresponds to the velocity field ${\bm v}$. The propagator functions and vertices are defined in Sec.~\ref{App1Z2}. 

Let us start with the graph $D_1$. Analytical expression for it reads
\begin{equation}
D_1=-2\pi B_0\int\frac{d\omega}{2\pi}\frac{d^d{\bm k}}{(2\pi)^d}\,
\delta(k_{\|})\, \frac{k_{\bot}^{5-d-(\xi+\eta)}}{\omega^2+\left(\alpha_0\nu_{\bot 0}k_{\bot}^{2-\eta}\right)^2}
\frac{p_{\|}(p_{\|}-k_{\|})}{-{i}\omega+\epsilon(p-k)}k,
\end{equation}
where $\epsilon(k)$ is denoted in Eq.~\eqref{lines3}, ${\bm p}$ is an external momenta, and $p_{\|}=({\bm p}\cdot {\bm n})$. Since $d_\Gamma=2$ for this function, we are looking for the term proportional to $p_{\|}^2$. 
Owing to this fact, after trivial integration over $k_{\|}$ and after integration over the frequency $\omega$ one obtains 
\begin{equation}
D_1=-B_0p_{\|}^2\frac{1}{2\alpha_0\nu_{\bot_0}^2}\int_{k_\bot>m} \frac{d^{d-1}{\bm k}_{\bot}}{(2\pi)^{d-1}}\frac{k_{\bot}^{5-d-(\xi+\eta)}}{k_{\bot}^{2-\eta}\left(\alpha_0k_{\bot}^{2-\eta}+k_{\bot}^{2}\right)}.
\end{equation}
The integration over the internal momenta ${\bm k_{\bot}}$ can be
simplified in the MS scheme, in which all the anomalous
dimensions are independent of the regularizators like $\xi$
and $\eta$. Hence, we may choose them arbitrary with the only restriction that
our diagrams have to remain UV finite~\cite{Ant3}. The most convenient way is to put $\eta=0$. Thus, after the integration one obtains
\begin{equation}
D_1=-p_{\|}^2\frac{B_0}{2\alpha_0(\alpha_0+1)\nu_{\bot_0}^2}\,\frac{S_{d-1}}{(2\pi)^{d-1}}\,\frac{m^{-\xi}}{\xi},
\label{ans1}
\end{equation}
where $S_{d-1}$ is the area of the unit sphere in the $(d-1)$-dimensional space.

The analytical expression for the graph $D_2$ reads
\begin{equation}
D_2=-D_0\int\frac{d\omega}{2\pi}\frac{d^d{\bm k}}{(2\pi)^d}\frac{p_{\|}(p_{\|}-k_{\|})}{[\omega^2+\epsilon^2(k)][-{i}\omega+\epsilon(p-k)]}.
\label{tmp1}
\end{equation}
Since now we do not have $\delta(k_{\|})$ in our integrand, expansion of expression~\eqref{tmp1} over $p_{\|}$ gives us two terms. 
First one referred to as $I_1$ reads
\begin{equation}
I_1=-D_0p_{\|}^2\int\frac{d\omega}{2\pi}\frac{d^d{\bm k}}{(2\pi)^d}\frac{1}{[\omega^2+\epsilon^2(k)][-{i}\omega+\epsilon(k)]}.
\end{equation}
After integration over $\omega$ one obtains
\begin{equation}
I_1=-p_{\|}^2\frac{D_0}{4}\int\frac{d^d{\bm k}}{(2\pi)^d}\frac{1}{\left(\nu_{\|_0}k^2_{\|}+\nu_{\bot_0}k^2_{\bot}\right)^2}.
\label{tmp2}
\end{equation}
To preform integration in Eq.~\eqref{tmp2} it is convenient to pass to the new variables $l_{\|}=\nu^{1/2}_{\|_0}k_{\|}$ and $l_{\bot}=\nu^{(d-1)/2}_{\bot_0}k_{\bot}$ which adsorb viscosity coefficients. Thus, expression for $I_1$ reads
\begin{equation}
I_1=-p_{\|}^2\frac{D_0}{4\nu^{1/2}_{\|_0}\nu^{(d-1)/2}_{\bot_0}}\int\frac{d^d{\bm l}}{(2\pi)^d}\frac{1}{l^4},
\end{equation}
where $l^2=l_{\|}^2+l_{\bot}^2$. Substituting the value of the logarithmic dimension $d=4-\varepsilon$ one finally obtains
\begin{equation}
I_1=-p_{\|}^2\frac{D_0}{4\nu^{1/2}_{\|_0}\nu^{3/2}_{\bot_0}}\frac{{S_d}}{(2\pi)^d}\int_{l>m}\frac{dl}{l^{1+\varepsilon}}=
-p_{\|}^2\frac{D_0}{4\nu^{1/2}_{\|_0}\nu^{3/2}_{\bot_0}}\frac{{S_d}}{(2\pi)^d}\,\frac{m^{-\varepsilon}}{\varepsilon}. 
\end{equation}

To write the second term in Eq.~\eqref{tmp1} referred to as $I_2$ we should use expansion
\begin{equation}
\frac{1}{\epsilon(k)+\epsilon(p-k)}=\frac{1}{2\epsilon(k)}\left[1+\frac{\nu_{\|_0}p_{\|}k_{\|}+\nu_{\bot_0}({\bm p}_{\bot}\cdot {\bm k}_{\bot})}{\epsilon(k)}\right]+O\left(p^2\right).
\end{equation}
Using the fact that only terms even in ${\bm k}$ give nonzero contributions one obtains
\begin{equation}
I_2=p_{\|}\frac{D_0}{4}\int\frac{d^d{\bm k}}{(2\pi)^d}\frac{k_{\|}\left[\nu_{\|_0}p_{\|}k_{\|}+\nu_{\bot_0}({\bm p}_{\bot}\cdot {\bm k}_{\bot})\right]}{\left(\nu_{\|_0}k^2_{\|}+\nu_{\bot_0}k^2_{\bot}\right)^3}.
\end{equation}
After the same replacement of the variables as the one we used in Eq.~\eqref{tmp2} expression for $I_2$ takes the form
\begin{equation}
I_2=p_{\|}\frac{D_0}{4\nu^{1/2}_{\|_0}\nu^{(d-1)/2}_{\bot_0}}\int\frac{d^d{\bm l}}{(2\pi)^d}\frac{p_{\|}l^2_{\|}+l_{\|}({\bm p}_{\bot}\cdot {\bm l}_{\bot})}{l^6}.
\label{tmp3}
\end{equation}
In order to integrate over the vector ${\bm l}$ we need to average our expression over the angles:
\begin{equation}
\int d{\bm l} f({\bm l})=
S_{d} \int_{l>m} dl\,l^{d-1}\,
\left\langle f({\bm l})\right\rangle,
\end{equation}
where $\langle\cdots\rangle$ is the averaging over the unit sphere in the
$d$-dimensional space. In particular case of two indices it reads
\begin{align}
\left\langle \frac{l_i l_j}{l^2} \right\rangle
&= \frac{\delta_{ij}}{d}.
\end{align}
For the second term in Eq.~\eqref{tmp3} this gives
\begin{equation}
\int\frac{d^d{\bm l}}{(2\pi)^d}\frac{l_{\|}({\bm p}_{\bot}\cdot {\bm l}_{\bot})}{l^6}=\int\frac{d^d{\bm l}}{(2\pi)^d}\frac{l_in_il_jp_{j\bot}}{l^6}=\frac{S_d}{d}\delta_{ij}n_ip_{j\bot}\int_{l>m} \frac{dl}{l^4}=0,
\end{equation}
where the last equality follows from the fact that $\delta_{ij}n_ip_{j\bot}=n_jp_{j\bot}=0$.

The first term in Eq.~\eqref{tmp3} is nonzero and after substitution $d=4-\varepsilon$ gives
\begin{equation}
I_2=p^2_{\|}\frac{D_0}{4\nu^{1/2}_{\|_0}\nu^{3/2}_{\bot_0}}\int\frac{d^d{\bm l}}{(2\pi)^d}\frac{n_in_jl_il_j}{l^6}=
p^2_{\|}\frac{D_0}{16\nu^{1/2}_{\|_0}\nu^{3/2}_{\bot_0}}\frac{{S_d}}{(2\pi)^d}\,\frac{m^{-\varepsilon}}{\varepsilon}. 
\end{equation}
Combining together expressions for $I_1$ and $I_2$ one finally obtains
\begin{equation}
D_2=-p^2_{\|}\frac{3}{16}\frac{D_0}{\nu^{1/2}_{\|_0}\nu^{3/2}_{\bot_0}}\frac{{S_d}}{(2\pi)^d}\,\frac{m^{-\varepsilon}}{\varepsilon}. 
\label{ans2}
\end{equation}

The one-loop approximation for the 1-irreducible Green function $\langle h'h \rangle_{1-ir}$ reads
\begin{equation} 
\label{Dyson}
\langle h'h \rangle_{1-ir}= {i}\omega -\nu_{\|_0} p^2_{\|_0} -\nu_{\bot_0} p^2_{\bot_0}+\Sigma, 
\end{equation}
where $\Sigma$ is the self-energy operator and is represented by the sum of the graphs $D_1$ and $D_2$. Combining this expression with~\eqref{ans1} and~\eqref{ans2} and taking into account definitions of the coupling constants [see Eqs.~\eqref{D0}] one immediately obtains
the renormalization constants $Z_{\nu_{\|}}$ and $Z_{\nu_{\bot}}$, see Eq.~\eqref{ZM1}.

%%%%%%%%%%%%%%%%%%%%%%%%%%%%%%%%%%%%%%%%%%%%%%%%%%%%%%%%%%%%%%%%%%%%%%%%%%%%%%%%%%%%%%%%%%%%%%%%%%%%%%%%%%%%%%%%%%%%%%%%%%%%%%%%%%%%%%%%%%%%%%%%%%%%%%%%%%%%%%%%%%%%%%%%
\section{Calculation details of Model~2}
\label{App2}
%%%%%%%%%%%%%%%%%%%%%%%%%%%%%%%%%%%%%%%%%%%%%%%%%%%%%%%%%%%%%%%%%%%%%%%%%%%%%%%%%%%%%%%%%%%%%%%%%%%%%%%%%%%%%%%%%%%%%%%%%%%%%%%%%%%%%%%%%%%%%%%%%%%%%%%%%%%%%%%%%%%%%%%%

This section contains detailed calculations of the diagrams defining the renormalization constants $Z_{\nu_\parallel}$, $Z_{h}$, and $Z_{v}$ (see Sec.~\ref{App2Z}).
Since we deal with three divergent Green functions in Model~2, namely $\langle h'h \rangle_{1-ir}$, $\langle h'hv \rangle_{1-ir}$, and $\langle h'hh \rangle_{1-ir}$, we have to calculate both two-tailed and three-tailed graphs. 

Let us start with two-tailed graphs which enter the same expansion for function $\langle h'h \rangle_{1-ir}$ as Eq.~\eqref{Dyson}:
\begin{equation} 
\label{Dyson2}
\langle h'h \rangle_{1-ir}= {i}\omega -\nu_{\|_0} p^2_{\|_0} -\nu_{\bot_0} p^2_{\bot_0}+\Sigma.
\end{equation}
The graphs $\widetilde{D}_1$ and $\widetilde{D}_2$, the sum of which represents the self-energy operator $\Sigma$ (here and below graphs with tilde denote graphs for Model~2), are depicted by the same figures as shown in Eqs.~\eqref{d1d2white}. Moreover, the only difference between two models in Feynman rules is in the expression for the  propagator $\langle hh \rangle_0$ (see Sec.~\ref{App2Z2}) which does not enter the expression for $\widetilde{D}_1$. Thus, 
\begin{equation}
\widetilde{D}_1=D_1=-p_{\|}^2\frac{B_0}{2\alpha_0(\alpha_0+1)\nu_{\bot_0}^2}\,\frac{S_{d-1}}{(2\pi)^{d-1}}\,\frac{m^{-\xi}}{\xi}.
\label{ans3}
\end{equation}

The analytical expression for the graph $\widetilde{D}_2$ reads
\begin{equation}
\widetilde{D}_2=-D_0\int\frac{d\omega}{2\pi}\frac{d^d{\bm k}}{(2\pi)^d}\frac{2\pi\delta(\omega)}{\epsilon^2(k)}\frac{p_{\|}(p_{\|}-k_{\|})}{-{i}\omega+\epsilon(p-k)}.
\label{tmp5}
\end{equation}
Integration of Eq.~\eqref{tmp5} over the frequency is trivial and gives
\begin{equation}
\widetilde{D}_2=-D_0\int\frac{d^d{\bm k}}{(2\pi)^d}\frac{1}{\epsilon^2(k)}\frac{p_{\|}(p_{\|}-k_{\|})}{\epsilon(p-k)}.
\end{equation}
The general logic of the integration of expression for $\widetilde{D}_2$ is absolutely the same as in the previous section for the graph $D_2$, see Eqs.~\eqref{tmp1}~-- \eqref{ans2}: we have to extract the term $O\left({\bm p}^2\right)$ from the expression $p_{\|}(p_{\|}-k_{\|})/\epsilon(p-k)$ and then perform integration over the vector ${\bm k}$ taking into account $\tilde\varepsilon=6-d$. Therefore, we will omit these algebraic steps and write directly the final result: 
\begin{equation}
\widetilde{D}_2=-p^2_{\|}\frac{2}{3}\frac{D_0}{\nu^{1/2}_{\|_0}\nu^{5/2}_{\bot_0}}\frac{{S_d}}{(2\pi)^d}\,\frac{m^{-\tilde\varepsilon}}{\tilde\varepsilon}. 
\label{ans4}
\end{equation}
By combining expressions~\eqref{ans3} and~\eqref{ans4} and substituting them into Eq.~\eqref{Dyson2} one immediately obtains the renormalization constants $Z_{\nu_{\|}}$ and $Z_{\nu_{\bot}}$, see Eqs.~\eqref{ZS2}.

Now let us consider the graphs that correspond to the function $\langle h'hh \rangle_{1-ir}$.  Using our Feynman rules we may construct six graphs: 
\begin{eqnarray}
\widetilde{D}_3=\raisebox{-3ex}{ \includegraphics[width=1.8truecm]{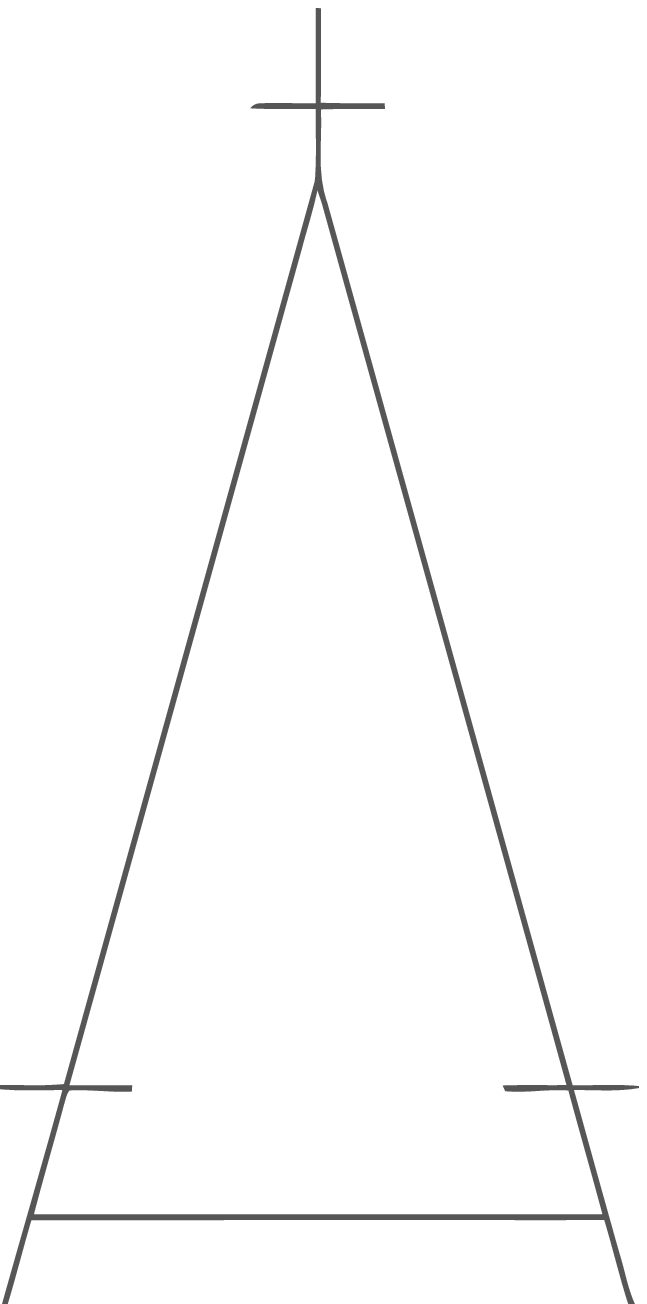}}, \quad \widetilde{D}_4=\raisebox{-3ex}{ \includegraphics[width=1.8truecm]{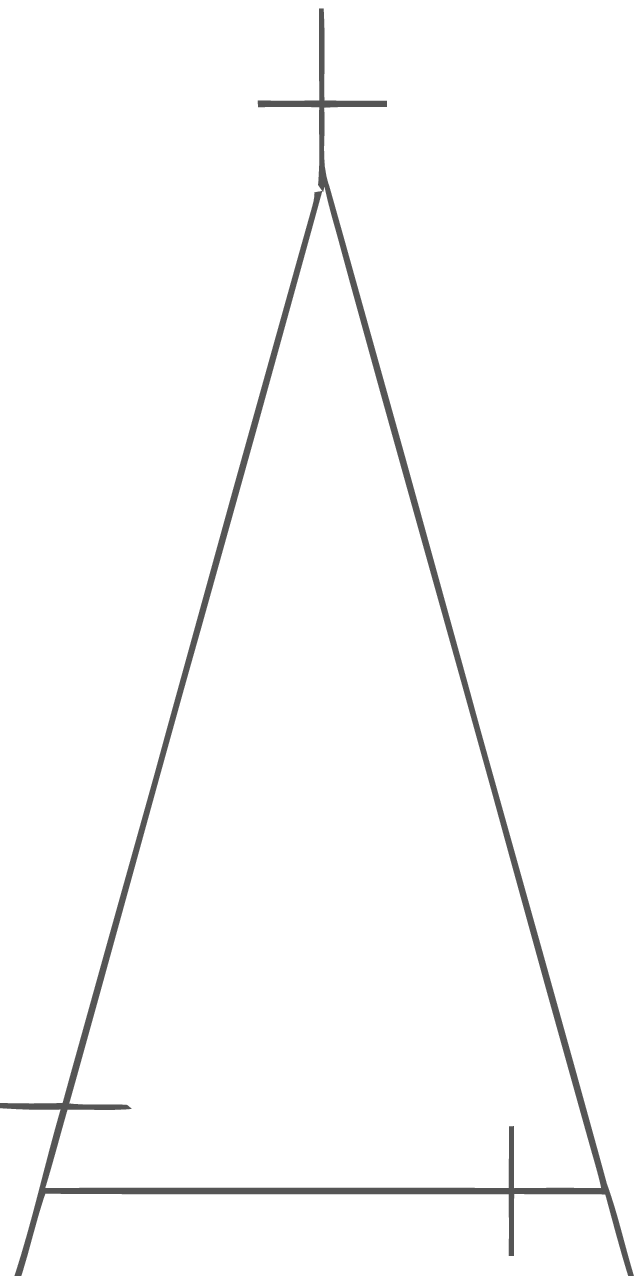}}, \quad \widetilde{D}_5=\raisebox{-3ex}{ \includegraphics[width=1.8truecm]{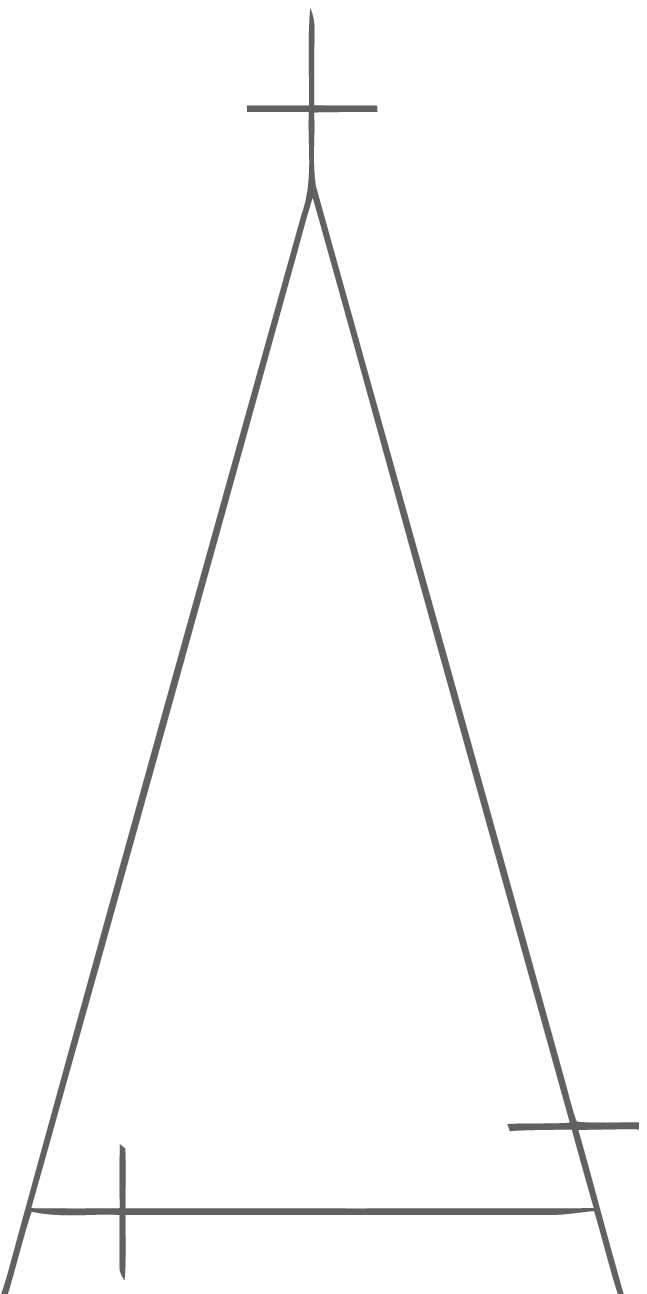}}\, , \nonumber
\end{eqnarray}
\begin{eqnarray}
\widetilde{D}_6=\raisebox{-3ex}{ \includegraphics[width=1.8truecm]{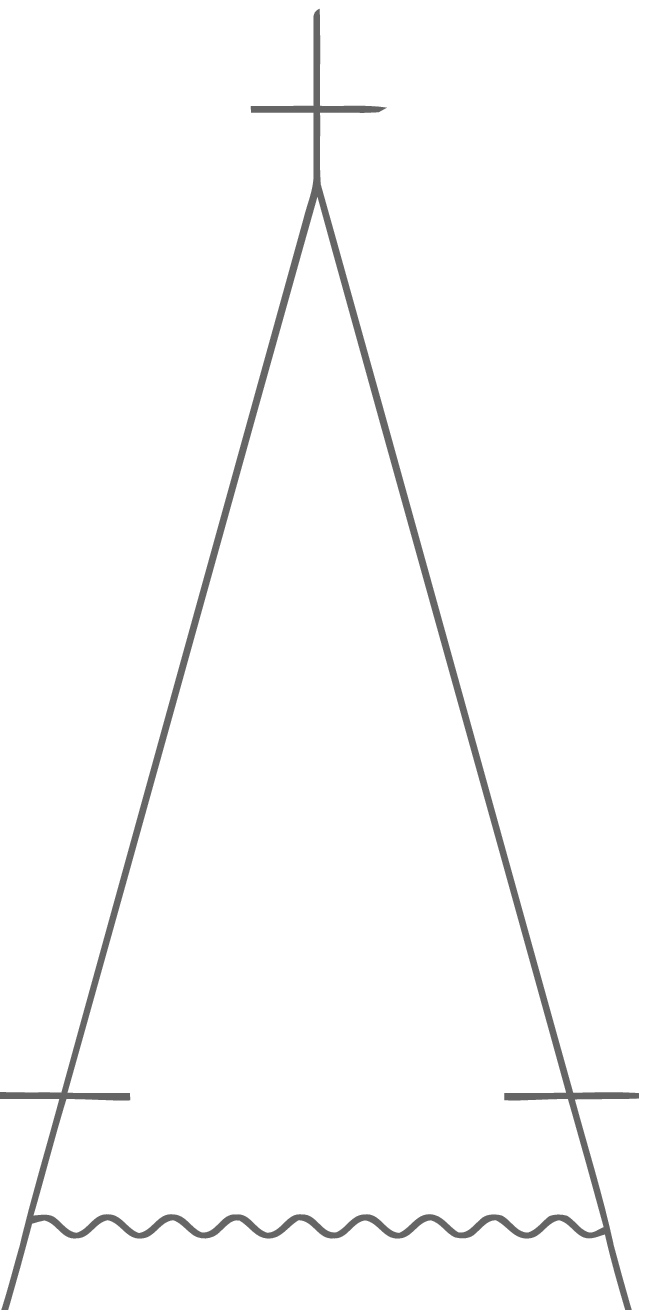}}, \quad
\widetilde{D}_7=\raisebox{-3ex}{ \includegraphics[width=1.8truecm]{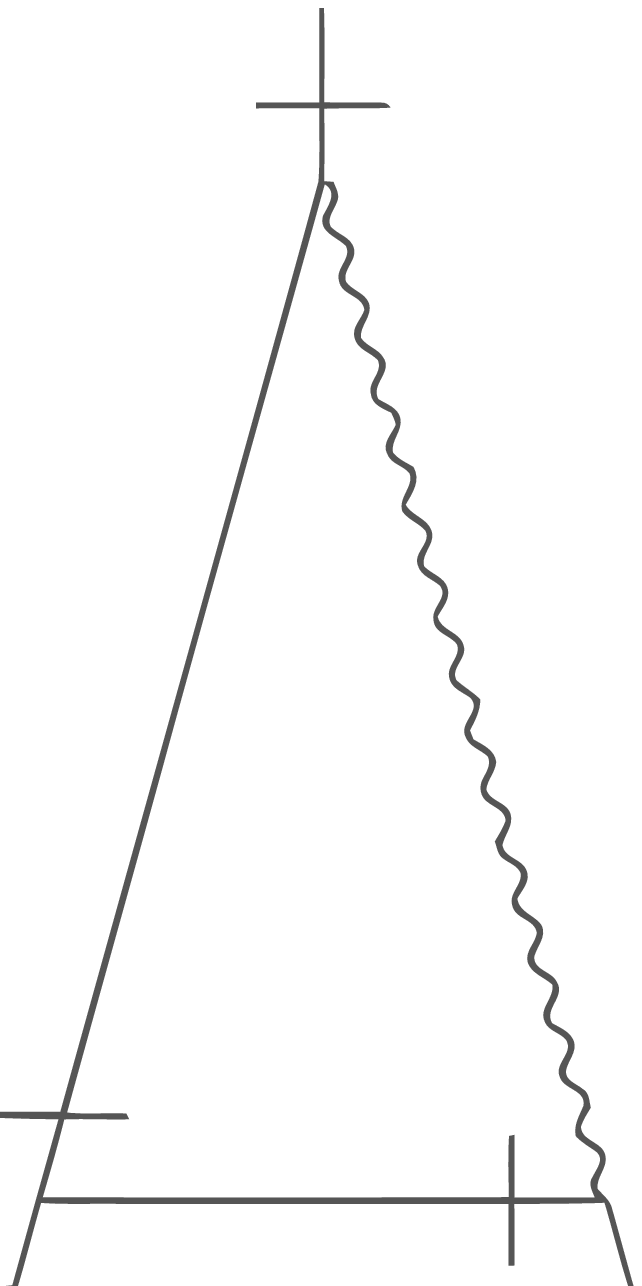}}  \quad \text{and} \quad \widetilde{D}_8=\raisebox{-3ex}{ \includegraphics[width=1.8truecm]{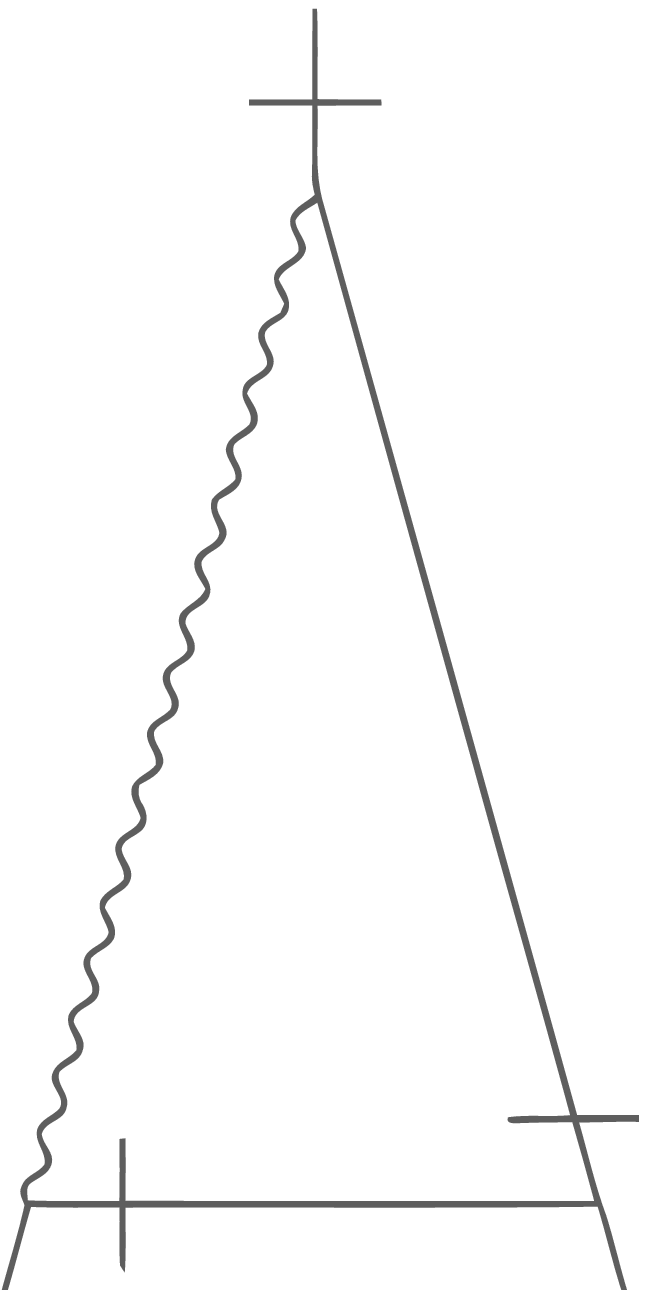}}\, .
\end{eqnarray}

Let us start with the last one, the graph $\widetilde{D}_8$. The analytical expression for it reads
\begin{equation}
\widetilde{D}_8=2\pi {i}^3B_0\int\frac{d\omega}{2\pi}\frac{d^d{\bm k}}{(2\pi)^d}\delta(k_{\|})
\frac{k_{\bot}^{5-d-(\xi+\eta)}}{\omega^2+\left(\alpha_0\nu_{\bot 0}k_{\bot}^{2-\eta}\right)^2}
\frac{(p_{\|}+q_{\|})(p_{\|}+q_{\|}+k_{\|})(k_{\|}+q_{\|})}{[-{i}\omega+\epsilon(p+k)][-{i}\omega+\epsilon(p+q+k)]}.
\label{tmp6}
\end{equation}
Here ${\bm p}$ and ${\bm q}$ are external momenta, ${\bm k}$ is momentum of integration. 
Since the divergence index for this graph $d_\Gamma=1$, we are looking for the terms proportional to ${\bm p}^1$ or ${\bm q}^1$. 
Expression~\eqref{tmp6} has a factor $(p_{\|}+q_{\|})$ from the very beginning, therefore, we may immediately put ${\bm p}={\bm q}=0$ in all the other coefficients. 
This observation together with $\delta(k_{\|})$ presented in r.h.s. of Eq.~\eqref{tmp6} leads directly to the fact that $\widetilde{D}_8=0$. 

The same feature is true also for the graphs $\widetilde{D}_6$ and $\widetilde{D}_7$. The fact that divergent parts of all three graphs containing velocity propagator $\langle vv \rangle_{0}$ are equal to zero leads to independence of renormalization constant $Z_h$ from the coupling constant $w$. 
Moreover, this effect holds true in all orders of perturbation theory: in any multiloop graph of such type we may choose the same direction of momenta flow as we chose in Eq.~\eqref{tmp6}. 

Despite the fact that we consider Green function $\langle h'hh \rangle_{1-ir}$ now, let us mention in this place that the function $\langle h'hv \rangle_{1-ir}$ does not have graphs similar to $\widetilde{D}_6$~-- $\widetilde{D}_8$. The reason for this is that we simply do not have a vertex with two fields ${\bm v}$ in our Feynman rules. Therefore, $Z_v$ also does not have dependence on the coupling constant $w$ in all orders of perturbation theory. 

This feature of the model has a great consequence for the critical dimensions at fixed points (3a) and (3b) which correspond to the case $g^*=0$, $w^*\neq0$.
Since $\gamma_v^*=\gamma_h^*=0$ in all orders of perturbation theory at these points, critical dimensions found in one-loop approximation [see Eqs.~\eqref{dim3a2}~-- \eqref{dim3b2}] are, in fact, exact. The situation is similar to Model~1. The difference is that we do not have any relation like Eq.~\eqref{relation} in Model~2 from which this fact would follow obviously; 
moreover, critical dimensions in Model~2 are exact only for points (3a) and (3b), for all other scaling regimes they have corrections in $\tilde\varepsilon$, $\xi$ and $\eta$ of second order and higher. 

Divergent parts of other three graphs are nonzero. 
The analytical expression for the graph $\widetilde{D}_3$ reads
\begin{equation}
\widetilde{D}_3=-{i}^3D_0\int\frac{d\omega}{2\pi}\frac{d^d{\bm k}}{(2\pi)^d}
\frac{2\pi\delta(\omega)}{\epsilon^2(k)}
\frac{(p_{\|}+q_{\|})(p_{\|}+q_{\|}+k_{\|})(k_{\|}+q_{\|})}{[-{i}\omega+\epsilon(p+k)][-{i}\omega+\epsilon(p+q+k)]}.
\end{equation}
After trivial integration over the frequency $\omega$, the term proportional to $(p_{\|}+q_{\|})$ takes the form
\begin{equation}
\widetilde{D}_3={i}(p_{\|}+q_{\|})D_0\int\frac{d^d{\bm k}}{(2\pi)^d}
\frac{k_{\|}^2}{\epsilon^4(k)}.
\label{tmp8}
\end{equation}
Using the same techniques as we described above, from Eq.~\eqref{tmp8} one immediately obtains 
\begin{equation}
\widetilde{D}_3={i}(p_{\|}+q_{\|})
\frac{1}{6}\frac{D_0}{\nu^{1/2}_{\|_0}\nu^{5/2}_{\bot_0}}\frac{{S_d}}{(2\pi)^d}\,\frac{m^{-\tilde\varepsilon}}{\tilde\varepsilon}. 
\label{tmp7}
\end{equation}
The graphs $\widetilde{D}_4$ and $\widetilde{D}_5$ are equal to each other and differ from $\widetilde{D}_3$ only by the sign. Thus, 
\begin{equation}
\widetilde{D}_4=\widetilde{D}_5=-{i}(p_{\|}+q_{\|})
\frac{1}{6}\frac{D_0}{\nu^{1/2}_{\|_0}\nu^{5/2}_{\bot_0}}\frac{{S_d}}{(2\pi)^d}\,\frac{m^{-\tilde\varepsilon}}{\tilde\varepsilon}. 
\end{equation}
The symmetry coefficients for the diagrams ${\widetilde D}_1$-- ${\widetilde D}_8$ are all equal 1.

The one-loop approximation for the 1-irreducible Green function $\langle h'hh \rangle_{1-ir}$ reads
\begin{equation} 
\label{Dyson3}
\langle h'hh \rangle_{1-ir}= V_{h'hh} + \widetilde{D}_3 + \widetilde{D}_4 + \widetilde{D}_5. 
\end{equation}

From Eqs.~\eqref{tmp7}~-- \eqref{Dyson3} one immediately arrives
at the renormalization constant $Z_h$, see Eqs.~\eqref{ZS2}. 
Since the graphs containing the propagator  $\langle vv \rangle_{0}$ vanish, 
the constant $Z_h$ coincides with its counterpart in the model without the turbulent environment~\cite{Stat}.

In the end of this section we turn to the graphs corresponding to the function $\langle h'hv \rangle_{1-ir}$. The only difference between them and graphs 
$\widetilde{D}_3$~-- $\widetilde{D}_5$ for the function $\langle h'hh \rangle_{1-ir}$ is the presence of the external field ${\bm v}$ instead of the field $h$. 
The cores of the graphs are the same. 
Thus, $Z_v=Z_h$.

%%%%%%%%%%%%%%%%%%%%%%%%%%%%%%%%%%%%%%%%%%%%%%%%%%%%%%%%%%%%%%%%%%1%%%%%%%%%%%%%%%%%%%%%%%%%%%%%%%%%%%%%%%%%%%%%%%%%%%%%%%%%%%%%%%%%%%%%%%%%%%%%%%%%%%%%%%%%%%%%%%%%%%%%%

\end{document}